\documentclass[12pt]{article}

\usepackage[T1]{fontenc}
\usepackage[section]{placeins}
\usepackage{rotating}
\usepackage{color}
\usepackage[english]{babel}
\usepackage{standalone}
\usepackage{float}
\usepackage{booktabs}
\usepackage{rotating}
\usepackage{graphics}
\graphicspath{ {Fig/} }
\usepackage{listings}
\usepackage{setspace}
\usepackage{natbib}
\bibpunct{(}{)}{;}{a}{,}{,}
\usepackage{url}
\usepackage[titletoc,title]{appendix}
\usepackage[left=1.5cm,right=1.5cm,top=2cm,bottom=2cm]{geometry}
\usepackage[font=footnotesize, labelfont={bf}, margin=1cm]{caption}
\usepackage{rotating}
\usepackage{array}
\usepackage{colortbl}
\usepackage{enumerate,multicol}
\usepackage{amsmath, amssymb, amsthm}
\usepackage{enumitem,bm}
\usepackage{parcolumns}

\usepackage[multiple]{footmisc}
\usepackage{authblk}
\usepackage[position=bottom]{subfig}
\usepackage{multirow}
\usepackage{makecell}
\usepackage{fancyvrb} 
\usepackage{listings,color} 
\usepackage{moreverb}
\usepackage{diagbox}
\usepackage{mathtools}
\usepackage{nccmath}
\usepackage{hyperref}

\hypersetup{hidelinks}

\usepackage{pifont}
\newcommand\ack{\section*{Acknowledgement}}

\makeatletter
\newcommand{\vast}{\bBigg@{7}}
\newcommand{\vastt}{\bBigg@{5}}
\makeatother

\begin{document}

\title{\bf A Bayesian Approach for Nonignorable Dropout in Bivariate Longitudinal Models}
\author[1]{Andrea Gabrio} 
\author[2]{Michael J. Daniels}
\author[3]{Gianluca Baio}
\affil[1]{\small \textit{Department of Methodology and Statistics, Faculty of Health Medicine and Life Sciences, Maastricht University, Peter Debeyeplein 1, Maastricht 6229 HA, NL.}}
\affil[2]{\small \textit{Department of Statistics, University of Florida, Gainesville, FL.}}
\affil[3]{\small \textit{Department of Statistical Science, University College London, 1-19 Torrington Place, London~WC1E~7HB,~UK.}}
\date{}
\maketitle

\begin{abstract}
Longitudinal data collected in clinical trials are almost always incomplete due to some of the participants dropping out from the study during the planned follow-up. A common strategy to handle nonresponse expresses missingness in terms of a dropout process, which is jointly analysed with the outcome process to facilitate the formulation of the missingness assumptions. However, when the outcome is multivariate, the identification of the dropout process becomes problematic, especially when individuals have different dropout times for each type of response, and sensitivity analysis is difficult. The modelling task may be also be complicated by data complexities (e.g.~skewness and spikes) which are difficult to capture through standard parametric methods. An example of this analysis framework occurs in trial-based economic evaluations, where a longitudinal bivariate response, formed by suitably-defined measures of effectiveness and costs, is analysed to inform policymakers about the cost-effectiveness of alternative interventions. 
We present a novel Bayesian nonparametric approach to handle a missing bivariate longitudinal outcome by jointly modelling the dropout process associated with each type of response while also taking into account the complexities of the data. We specify a flexible nonparametric model for the observed data and partially identify the distribution of the missing data with identifying restrictions conditional on the dropout indicators and sensitivity parameters. We explore alternative nonignorable scenarios through different priors for the sensitivity parameters. Our approach is motivated by, and applied to, data from a trial assessing the cost-effectiveness of a new treatment for intellectual disability and challenging~behaviour.
\end{abstract}



\section{Introduction}\label{intro}
The statistical analysis of health economic data has become an increasingly important component of clinical trials as it provides one of the earliest opportunities to assess the cost-effectiveness of alternative treatments and to guide the decisions of policymakers~\citep{glick2014economic}. The typical analysis is based on a bivariate outcome process formed by suitable measures of effectiveness and costs repeatedly collected at different times and then combined into aggregated measures of cost-effectiveness, e.g.~Quality Adjusted Life Years (QALYs). The main objective of the analysis is: a) to estimate population average effectiveness and costs in order to determine the most cost-effective intervention; and b) to assess the impact of the uncertainty in the model inputs on the decision-making process~\citep{Claxton}.

Individual level cost-effectiveness data are often characterised by a series of complexities which make the modelling task particularly challenging. First, the analysis of a bivariate outcome requires the use of methods that deal with correlation~\citep{OHaganb}. Second, data may exhibit "spikes" in the underlying distribution that may be difficult to capture using standard parametric models~\citep{Baio2014,Gabriob}. Finally, the recorded outcome process is often incomplete due to individuals dropping out or being observed intermittently during the follow-up, causing some observations to be~missing. 

According to recent reviews~\citep{Gabrio,Leurent2018}, standard practice in trial-based cost-effectiveness analysis uses only data from the individuals who are observed at each time point in the trial (completers). This, however, is an inefficient approach as the information from the responses of all partially-observed subjects is discarded and it is also likely biased unless the completers are a random sample of the individuals on each arm~\citep{Little2002}. When missingness depends on the unobserved values, the dropout process is nonignorable and must be explicitly modelled to draw valid inferences~\citep{Rubina}. Methodology to deal with nonignorable missingness is challenging as any analysis requires untestable assumptions and results are often sensitive to the particular assumptions made~\citep{Molenberghs1997}. Thus, it is desirable to assess the robustness of the inferences by varying these assumptions in a principled manner~\citep{Daniels2008}. 

\subsection{Joint modelling of partially-observed longitudinal data}\label{jlm}
We now recall some key results from the missing data literature~\citep{Little2002}. Let $\bm Y_i=(Y_{i1},\ldots,Y_{iJ})$ denote the vector of responses, which is (potentially) measured on~$i=1,\ldots,N$ individuals at $j=1,\ldots,J$ time points. Let $R_{ij}$ be an indicator variable telling whether $Y_{ij}$ is observed or missing for the $i$-th individual at measurement time $j$, that is $R_{ij}=\mathbb{I}(Y_{ij} \; \text{is observed})$. Given the response pattern $\bm R=(R_{i1},\ldots,R_{iJ})$, we denote the missing and observed variables in pattern $\bm R=\bm r$ as $\bm Y_{\bar{\bm r}}=(Y_{ij}:r_{ij}=0)$ and $\bm Y_{\bm r}=(Y_{ij}:r_{ij}=1)$,~respectively. 

For inference, we define a joint probability model for the response and missingness as $p(\bm y, \bm r \mid \bm \omega),$ parameterised by $\bm \omega$. We can factorise this joint model in terms of the product of the model for the response $p(\bm y \mid \bm \theta)$ indexed by $\bm \theta$, and the model for the missingness indicator $p(\bm r \mid \bm y, \bm \psi)$, often referred to as the \textit{missingness mechanism} and indexed by $ \bm \psi$, where $\bm \omega=(\bm \theta, \bm \psi)$.
The missingness mechanism is said to be \textit{ignorable} if the information in $\bm r$ can be completely ignored when making inference about $\bm \theta$, that is when the following three conditions hold: 
\begin{itemize}
\item[1)] $p(\bm r \mid \bm y, \bm \psi)=p(\bm r \mid \bm y_{\bm r}, \bm \psi)$, that is missingness only depends on the observed responses, so it is \textit{missing at random} (MAR);
\item[2)] the parameters $\bm \omega$ can be decomposed into \textit{distinct subsets} indexing the response model and the missingness mechanism, i.e.~$\bm \omega=(\bm \theta, \bm \psi)$;
\item[3)] for Bayesian inference, the parameters of the response model and the missing data mechanism are \textit{a priori independent}, that is $p(\bm \theta,\bm \psi)=p(\bm \theta)p(\bm \psi)$. 
\end{itemize}
Under conditions weaker than MAR, it may become reasonable to discard only a part of the information in the missing data indicators $\bm R$ when making inference about $\bm \theta$. Let $d(\bm R)$ denote a summary function of $\bm R$ and suppose we factorise the missingness mechanism as 
\begin{equation*}
p(\bm r \mid \bm y, \bm \psi)= p(d(\bm r) \mid \bm y, \bm \lambda)p(\bm r \mid \bm y, d(\bm r), \bm  \kappa),
\end{equation*}
where $\bm \lambda$ and $\bm \kappa$ are the two subsets of $\bm \psi$ indexing the model of $d(\bm R)$ and $\left[ \bm R \mid d(\bm R)\right]$, respectively. The missingness mechanism is said to be \textit{partially missing at random} given $d(\bm R)$ if $p(\bm r \mid \bm y, d(\bm r), \bm  \kappa)=p(\bm r \mid \bm y_{\bm r}, d(\bm r), \bm  \kappa)$. In addition, if $\bm \kappa$ is distinct from $(\bm \theta, \bm \lambda)$, all the information in $\bm R$ other than $d(\bm R)$ may be ignored and inferences about $\bm \theta$ will only require the specification of the joint model $p(\bm y, d(\bm r))$. This condition is known as \textit{partial ignorability}~\citep{Harel2009}. 

The choice of $d(\bm R)$ is typically based on those aspects of $\bm R$ that may be related to the missing values and are therefore worth modelling. For example, an intuitive choice of $d(\bm R)$ in longitudinal studies is the time of dropout of the subjects. Indeed, dropout can be seen as a two-stage selection process where in the first stage the time at which subjects drop out is determined, while in the second stage different patterns of missingness before the time of dropout (referred to as \textit{intermittent} missingness) are determined given the dropout time. If the second stage does not depend on any missing values, then the mechanism associated with this stage does not need to be explicitly specified and inferences about $\bm \theta$ will only require a joint model~$p(\bm y, d(\bm r))$.   

When neither MAR nor partial MAR are deemed plausible, missingness is referred to as \textit{nonignorable} and the joint model $p(\bm y,\bm r)$ requires untestable assumptions about the missing data. Often, this is due to the fact that $p(\bm r \mid \bm y, \bm \psi)\neq p(\bm r \mid \bm y_{\bm r}, \bm \psi)$, where missingness is \textit{missing not at random} (MNAR). This can be intuitively seen from the so called \textit{extrapolation factorisation}~\citep{Daniels2008}, which expresses the joint model for $\bm Y$ and $\bm R$ as
\begin{equation}\label{extfact}
p(\bm y, \bm r \mid \bm \omega)=p(\bm y_{\bar{\bm r}} \mid \bm y_{\bm r}, \bm r, \bm \theta_{E})p(\bm y_{\bm r}, \bm r \mid \bm \theta_{O}),
\end{equation}
where the parameters $\bm \theta_{O}$ index the \textit{observed data distribution} which can be identified from the observed data, while the parameters $\bm \theta_{E}$ index the \textit{extrapolation distribution} which cannot be identified from the data. Different approaches can be used to model $p(\bm y, \bm r, \mid \bm \omega)$ based on alternative factorisations of the joint distribution $p(\bm y, \bm r)$. One approach is given by \textit{pattern mixture models}~\citep{Little1992}, which factor the joint into the product of the marginal distribution $p(\bm r \mid \bm \pi)$, where $\bm r$ is typically referred to as the \textit{missingness pattern}, and the conditional distribution of the response given the pattern $p(\bm y \mid \bm r, \bm \phi)$. A key feature of pattern mixture models is that the extrapolation distribution appears explicitly in the model specification. This leads to a straightforward understanding of the impact that missing data assumptions have in the model and makes the specification of $\bm \theta_E$ more intuitive through sensitivity parameters~\citep{daniels2000}.

While different examples of joint modelling approaches for $\bm Y$ and $\bm R$ have been explored in applied works for univariate responses, the implementation of these methods for a bivariate response process has generally received little attention in the literature~\citep{Gaskins2016}, especially in an health economic analysis framework. Only recently,~\citet{gabrio2020bayesian} and~\citet{mason2021flexible} proposed multivariate parametric approaches to handle the complexities of the data while exploring different missingness assumptions, while~\citet{oganisian2020bayesian} explored the use of a Bayesian nonparametric approach to model the joint distribution of cost and survival time data assuming ignorability of the missingness mechanism. In this article, we improve current practice and present a novel Bayesian nonparametric model for conducting inference on a bivariate longitudinal outcome, which accommodates a sensitivity analysis to nonignorable missingness assumptions. Our work is partially motivated by the economic analysis of data from a randomised trial in people with intellectual disabilities and challenging~behaviour.

\subsection{Outline}\label{outline}
In Section~\ref{pbs} we provide details on the motivating data. Section~\ref{framework} introduces our proposed bivariate longitudinal model for dropout as well as the strategy to specify the model for the observed data distribution and to identify the extrapolation distribution. In Section~\ref{mnar} we discuss the approach to deal with nonignorability, including the role of sensitivity parameters, specification of their distributions, and estimation of the targeted quantities in the analysis. Section~\ref{results} reports information about the implementation and assessment of the model and presents the posterior results based on the motivating data, while Section~\ref{evaluation} summarises the health economic results. Finally, we provide some concluding remarks in Section~\ref{discussion}.

\section{The PBS trial}\label{pbs}
The Positive Behaviour Support (PBS) study~\citep{Hassiotis2018} is a multicenter randomised controlled trial that, among its objectives, aimed at evaluating the cost-effectiveness of a new intervention (PBS, $108$ individuals) relative to treatment as usual (TAU, $136$ individuals) for people suffering from intellectual disability and challenging behaviour. The primary instruments used to collect outcome data were the EQ-5D questionnaire~\citep{EQ5D} and family/paid carer clinic records for the effectiveness and cost measures, respectively. The health outcome is expressed in terms of utility scores, defined on the interval $[-0.594,1]$ where $1$ represents the perfect health state while negative values indicate states that are considered ``worse than death''. Costs were obtained from the clinic records and are expressed in~$\pounds{}$. Measurements were scheduled to be collected at baseline and $6$ and $12$ months~follow-up ($J=3$).

Let $\bm y_{ij}=(u_{ij},c_{ij})$ denote the bivariate outcome of interest and let $\bm u_i=(u_{i1},\ldots,u_{iJ})$ and $\bm c_i=(c_{i1},\ldots,c_{iJ})$ denote the vectors of utility and cost outcomes that were planned to be collected for individual $i$ at measurement time $j$, with~$j~\in~\{1,2,3\}$. Both outcomes were partially observed for some participants in the study. We group the individuals according to the missingness patterns and denote with $\bm r_{ij}=(r^u_{ij},r^c_{ij})$ a pair of indicator variables that take value $1$ if the corresponding outcome for individual $i$ at time $j$ is observed and $0$ otherwise. Finally, we denote the missing utility and cost dropout indicators with $d^u_i$ and $d^c_i$, which indicate the time individual $i$ has his/her last observed utility and cost measurement, respectively (dropout time). Table~\ref{patterns} shows the missingness patterns and dropout times for the control $(t=1)$ and intervention ($t=2$) group as well as the number of individuals and the observed mean responses within each~pattern in the PBS trial. 
\begin{table}
\centering
\scalebox{0.75}{
\begin{tabular}{c|cccccccc|ccccccccc|}
\toprule
 \multicolumn{1}{c}{}  & &  \multicolumn{5}{c}{\textbf{control} ($t=1$)} & & & &\multicolumn{5}{c}{\textbf{intervention} ($t=2$)}&\multicolumn{1}{c}{} &  \multicolumn{1}{c}{}\\
\multicolumn{1}{c}{}   & $u_1$ & $c_1$ & $u_2$ & $c_2$ & $u_3$ & $c_3$ & $d^u;d^c$ & $n_{\bm r 1}$ & $u_1$ & $c_1$ & $u_2$ & $c_2$ & $u_3$ &  \multicolumn{1}{c}{$c_3$} &  \multicolumn{1}{c}{$d^u;d^c$} &  \multicolumn{1}{c}{$n_{\bm r 2}$} \\
\midrule
$\bm r =\bm 1 $  & 1 & 1 & 1 & 1 & 1 & 1 & \multirow{2}{*}{3;\;3} & \multirow{2}{*}{108} & 1 & 1 & 1 & 1 & 1& 1 & \multirow{2}{*}{3;\;3} &  \multirow{2}{*}{96}\\ 
mean   & 0.678 & 1546 & 0.684 & 1527 & 0.680 & 1520 & &  & 0.726 & 2818 & 0.771 & 2833 & 0.759 & 2878 & & \\[1ex]
$\bm r$  & 0 & 1 & 1 & 1 & 1 & 1 & \multirow{2}{*}{3;\;3} & \multirow{2}{*}{7} & 0 & 1 & 1 & 1 & 1& 1& \multirow{2}{*}{3;\;3} & \multirow{2}{*}{5}\\ 
mean   & -- & 1310 & 0.704 & 1440 & 0.644 & 1858 & &  & -- & 2573 & 0.780 & 2939 & 0.849 & 2113 & & \\ [1ex]
$\bm r$  & 1 & 1 & 0 & 1 & 1 & 1 & \multirow{2}{*}{3;\;3} & \multirow{2}{*}{4} & 1 & 1 & 0 & 1 & 1& 1& \multirow{2}{*}{3;\;3} & \multirow{2}{*}{1}\\ 
mean   & 0.709 & 1620 & -- & 1087 & 0.737 & 851 & & & 0.467 & 9649 & -- & 4828 & 0.259 & 4930 & & \\ [1ex]
$\bm r$   & 1 & 1 & 1 & 1 & 0 & 1 & \multirow{2}{*}{2;\;3} & \multirow{2}{*}{2} & 1 & 1 & 1 & 1 & 0 & 1& \multirow{2}{*}{2;\;3} & \multirow{2}{*}{1}\\ 
mean   & 0.564 & 640 & 0.648 & 512 & -- & 286 &  & & 0.817 & 3788 & 0.884 & 0 & -- & 0 & & \\ [1ex]
$\bm r$   & 1 & 1 & 0 & 0 & 1 & 1 & \multirow{2}{*}{3;\;3} & \multirow{2}{*}{4} & 1 & 1 & 0 & 0 & 1 & 1 & \multirow{2}{*}{3;\;3} & \multirow{2}{*}{1}\\ 
mean   & 0.716 & 2834 & -- & -- & 0.634 & 679 &  & & 0.501 & 3608 & -- & -- & 0.872 & 4781 & &\\ [1ex]
$\bm r$  & 1 & 1 & 0 & 0 & 0 & 0 & \multirow{2}{*}{1;\;1} & \multirow{2}{*}{4} & 1 & 1 & 0 & 0 & 0 & 0 & \multirow{2}{*}{1;\;1} & \multirow{2}{*}{4}\\ 
mean   & 0.434 & 1528 & -- & -- & -- & -- &  & & 0.760 & 3086 & -- & -- & -- & -- & & \\ [1ex]
$\bm r$   & 0 & 1 & 0 & 1 & 1 & 1 & \multirow{2}{*}{3;\;3} & \multirow{2}{*}{2} & 0 & 1 & 0 & 1 & 1& 1& \multirow{2}{*}{-;\;-} & \multirow{2}{*}{0}\\ 
mean   & -- & 595 & -- & 397 & 0.483 & 69 &  & & -- & -- & -- & -- & -- & -- & & \\ [1ex]
$\bm r$   & 1 & 1 & 1 & 1 & 0 & 0 & \multirow{2}{*}{2;\;2} & \multirow{2}{*}{2} & 1 & 1 & 1 & 1 & 0 & 0 & \multirow{2}{*}{-;\;-} & \multirow{2}{*}{0}\\ 
mean  & 0.743 & 1434 & 0.705 & 1606 & -- & -- & & & -- & -- & -- & -- & -- & -- & & \\ [1ex]
$\bm r$  & 1 & 1 & 0 & 1 & 0 & 1 & \multirow{2}{*}{1;\;3} & \multirow{2}{*}{3} & 1 & 1 & 0 & 1 & 0 & 1 & \multirow{2}{*}{-;\;-} & \multirow{2}{*}{0}\\ 
mean  & 0.726 & 1510 & -- & 432 & -- & 976 & &  & -- & -- & -- & -- & -- & -- & & \\ 
\bottomrule
\end{tabular}
}\caption{\label{patterns}Missingness patterns for $\bm y_j=(u_j,c_j)$ in the PBS study, with $j=1,2,3$. For each pattern and treatment group, the dropout indicators $(d^u;d^c)$, the number of subjects ($n_{r t}$) and the observed mean outcomes are reported. The absence of response values or individuals within a pattern is denoted~with~--.}
\end{table}
The number of observed patterns is relatively small and baseline costs are the only fully observed variables. With the exception of the dropout patterns where $d^u=d^c$, all other observed patterns have a relatively small number of individuals.

\begin{figure}[!h]
\centering
\subfloat[control]{\includegraphics[scale=0.46]{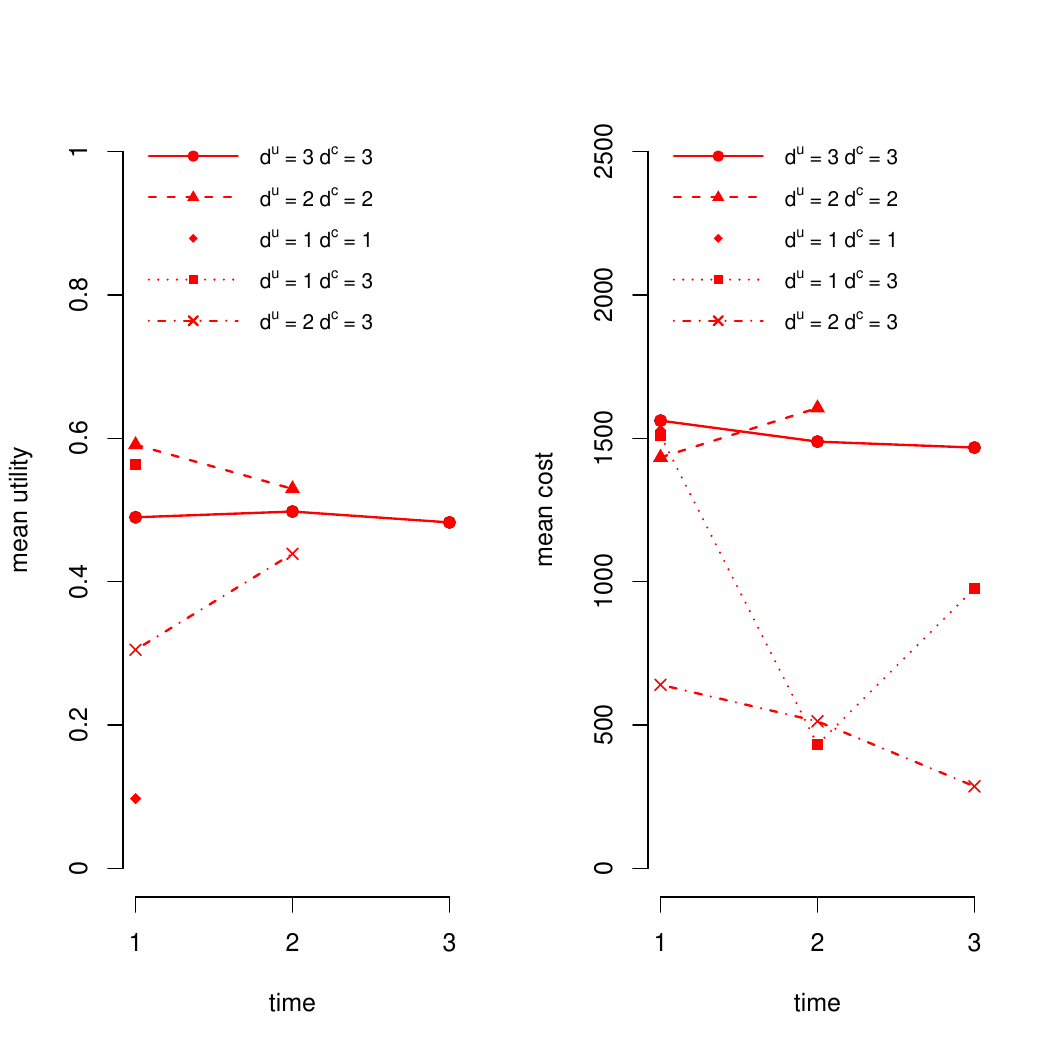}}
\subfloat[intervention]{\includegraphics[scale=0.46]{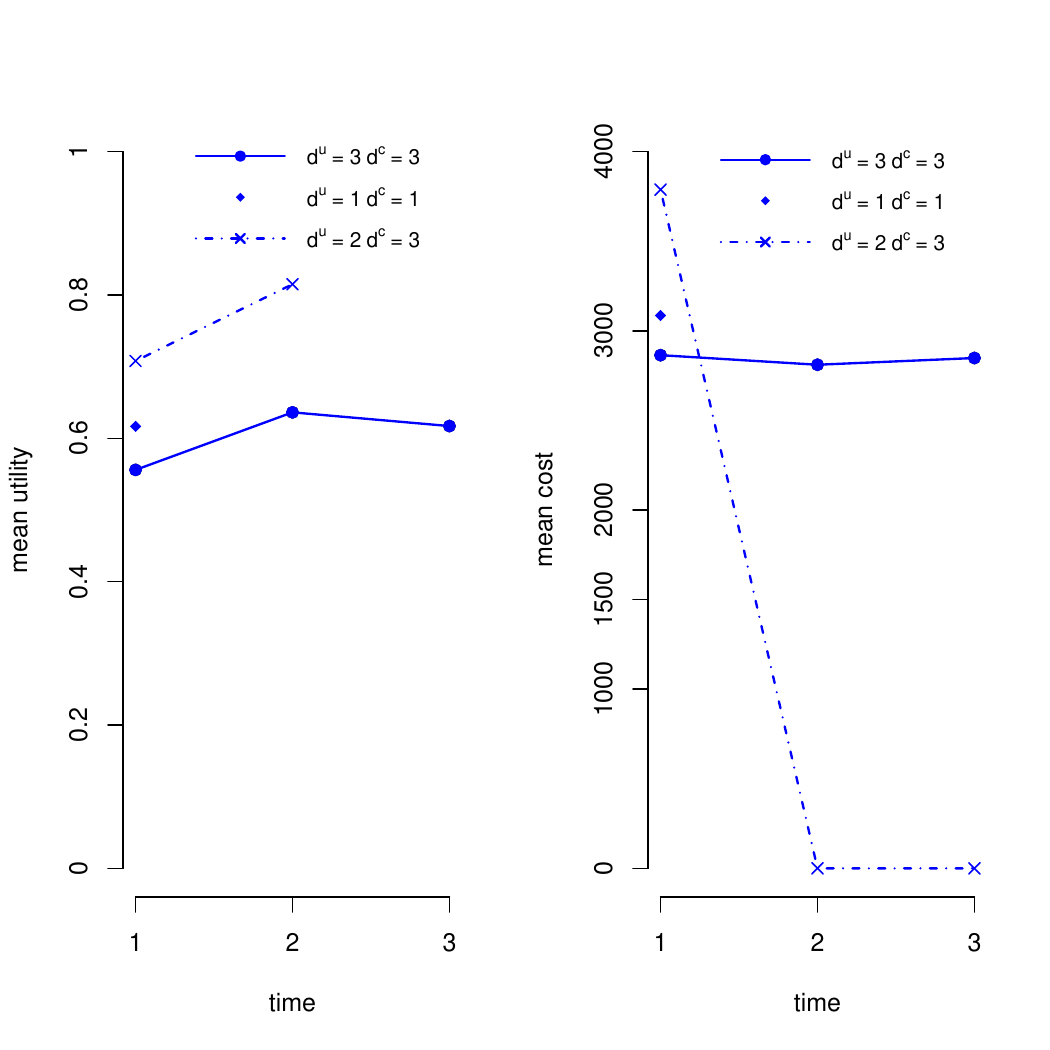}}
\caption{Empirical mean profiles for different utility and cost dropout times ($1=\text{baseline},2=\text{6-months},3=\text{12-months}$) in the control (panel a) and intervention (panel b) group of the PBS study.}
\label{fig_pat_drop}
\end{figure}

Figure~\ref{fig_pat_drop} shows the mean profiles associated with each observed dropout pattern $\bm d=(d^u,d^c)$, separately for the two treatment groups of the study.
People with different dropout times for $d^u$ and $d^c$ have considerable variations between their observed patterns in both treatments. For example: the completers ($d^u=d^c=3$) show a relatively stable mean profile for both costs and utilities; individuals who dropped out at $6$ months ($d^u=d^c=2$) show a decline in mean utilities and an increase in mean costs with respect to their mean baseline values; individuals who dropped out at $6$ months only for the utilities but who completed the study for costs ($d^u=2, d^c=3$) show an increase in mean utilities with respect to their mean baseline value and a systematic decline in mean costs over time.

In the next section, the general modelling framework for handling a partially-observed bivariate longitudinal outcome is illustrated. First, we provide our formulation of partial ignorability assumptions and show how it can be used to support the specification of a longitudinal bivariate model for dropout. Next, we present our modelling approach for the observed data and the identification strategy proposed to identify the unidentified outcome distributions for the non-completers.

\section{Modelling Framework}\label{framework}
We propose a strategy to model the joint distribution of the bivariate longitudinal outcome $\bm y=(\bm u, \bm c)$ and missingness indicators $\bm r=(\bm r^u, \bm r^c)$ under partial-ignorability conditional on the dropout indicators $\bm d=(d^u, d^c)$. Our modelling framework is based on three steps. First, we specify a model for the joint distribution using a pattern mixture approach: $p(\bm y, \bm r \mid \bm \omega)=p(\bm r\mid \bm \pi)p(\bm y \mid \bm r, \bm \phi)$, where $\bm \omega=(\bm \pi, \bm \phi)$. Second, using the extrapolation factorisation (Equation~\ref{extfact}), we retrieve the observed data distribution $p(\bm y_{\bm r}, \bm r \mid \bm \theta_{O})$ and leave the extrapolation distribution unidentified. Third, we identify $p(\bm y_{\bar{\bm r}} \mid \bm y_{\bm r}, \bm r, \bm \theta_{E})$ assuming partial-ignorability conditional on the dropout indicators $\bm d\,$ and through a combination of identifying restrictions and sensitivity parameters. We initially fit the model under a benchmark scenario and then assess the robustness of the results to alternative missingness assumptions by using different informative prior distributions on the sensitivity~parameters.

The quantities of interest for the economic assessment are the posterior distributions of the marginal mean QALYs and total costs $\bm \mu_{t}=(\mu_{et},\mu_{ct})$ in each treatment group. These are computed as linear combinations of the marginal mean utility ($\mu^u_{jt}$) and costs ($\mu^c_{jt}$) at each time in the study using standardised formulae~\citep{Drummond}. 

\subsection{Partial Ignorability for a Bivariate Longitudinal Outcome}\label{pi_biv}
Although partial ignorability based on a dropout indicator $d$ is easily defined for univariate longitudinal outcomes, its formulation becomes less straightforward when dealing with a bivariate or, in general, multivariate outcome process, since individuals may be associated with dropout times that differ between outcomes. To address this problem we propose to identify the joint model $p(\bm r, \bm y)$ using a summary function of the missingness indicators $d(\bm r^u, \bm r^c)$ that can facilitate the formulation of partial ignorability assumptions for a longitudinal bivariate outcome. 

More specifically, for each individual, we define two dropout indicators $d^{max}=\text{max}(d^u,d^c)$ and $d^{min}=\text{min}(d^u,d^c)$. For example, individuals who have completed data for both outcomes, $d(\bm r^u=\bm 1, \bm r^c=\bm 1)$, are associated with $d^{max}=d^{min}=J$, while individuals who have completed data only for costs but not for utilities, $d(\bm r^u \neq \bm 1, \bm r^c=\bm 1)$, are associated with $d^{min}<J, d^{max}=J$. Next, we use $d^{min}$ and $d^{max}$ to define partial ignorability such that, conditional on these indicators, inference can be made on the joint model $p(\bm y, \bm r)$ by ignoring all information in $\bm r$ which is not contained in $\bm d$. Different values for $d^{min}$ and $d^{max}$ have different implications with respect to which distributions can be identified from the observed data, so that the identification of $p(\bm y, \bm r)$ varies according to whether $d^{min}=d^{max}$ or~$d^{min}~\neq~d^{max}$.

\subsection{Observed Data and Extrapolation Distributions}\label{model_dropout}
We can factor the joint distribution of $\bm y$ and $\bm r$ using the extrapolation factorisation:
\begin{equation*}
p(\bm y, \bm r \mid \bm \omega)=p(\bm y_{\bar{\bm r}} \mid \bm y_{\bm r}, \bm r, \bm \omega)p(\bm y_{\bm r}, \bm r \mid  \bm \omega)
\end{equation*}
where $\bm y_{\bm r}$ and $\bm y_{\bar{\bm r}}$ indicate the observed and missing outcome according to the values of $\bm r$. To extract the observed data distribution $p(\bm y_{\bm r}, \bm r \mid  \bm \omega)$, we use a working model $p^{\star}$ for the joint distribution of the outcome and missingness~\citep{Linero2015}. Essentially, the idea is to use the working model $p^{\star}(\bm y, \bm r \mid  \bm \omega)$ to draw inferences about the distribution of the observed data $p(\bm y_{\bm r}, \bm r \mid \bm \omega)$ by integrating out the missing responses:
\begin{equation}\label{wm}
p(\bm y_{\bm r}, \bm r \mid \bm \omega)=\int p^{\star}(\bm y, \bm r \mid  \bm \omega)d \bm y_{\bar{\bm r}}.
\end{equation}
This approach avoids direct specification of the joint distribution of the observed and missing data, which has the undesirable consequence of identifying the extrapolation distribution with assumptions that are difficult to check. Indeed, since we use $p^{\star}(\bm y, \bm r \mid  \bm \omega)$ only to obtain a model for $p(\bm y_{\bm r}, \bm r \mid  \bm \omega)$ and not as a basis for inference, the extrapolation distribution is left unidentified. Any inference depending on the observed data distribution may be obtained using the working model as the true model, with the advantage that it is often easier to specify a model for the the full data $p(\bm y,\bm r)$ compared with a model for the observed data~$p(\bm y_{\bm r},\bm r)$. 

We specify $p^\star$ using a pattern mixture approach, factoring the joint $p(\bm y,\bm r \mid \bm \omega)$ as the product of the marginal distribution of the missingness patterns $p(\bm r \mid \bm \pi)$ and the distribution of the outcome conditional on the patterns $p(\bm y\mid \bm r,\bm \theta)$, respectively indexed by the distinct parameter vectors $\bm \pi$ and $\bm \theta$. We then express the missingness through the dropout indicators $d(\bm r)=\bm d=(d^{min},d^{max})$. The model is specified in terms of the full data $\bm y$, formed by the observed $\bm y_{\bm r}$ and missing $\bm y_{\bar{\bm r}}$ values, either intermittent or due to dropout.
Three components form the observed data distribution:
\begin{itemize}
\item[1] The distribution of the dropout indicators $p(\bm r \mid \bm \pi)$ and the outcome distributions of the completers $p(\bm y \mid \bm r = \bm 1, d^{min}=J, d^{max}=J)$, which are fully identified from the data. 
\item[2] The outcome distributions among the non-completers with $d^{min}=d^{max}$: 
\begin{equation*}
p(\bm y_{\bm r} \mid \bm r \neq \bm 1, d^{min}=d^{max}) = \prod_{j=1}^{d^{min}}p(\bm y_{\bm r j} \mid \bar{\bm y}_{\bm r j-1}, \bm r \neq \bm 1, d^{min}=d^{max}),
\end{equation*}
where $\bm y_{\bm rj}$ and $\bar{\bm y}_{\bm r j-1}$ are the observed outcomes at time $j$ and the observed history of the outcomes up to $j-1$, respectively. These distributions are identified under partial ignorability for $d^{min}$ and~$d^{max}$. The observed data part is obtained by integrating out all intermittent missing values~$\bm y_{\bar{\bm r}j}$ for $j \leq d^{min}$.
\item[3] The outcome distributions among the non-completers with $d^{min} \neq d^{max}$: 
\begin{align*}
p(\bm y_{\bm r} \mid \bm r \neq \bm 1, d^{min} \neq d^{max}) & = \prod_{j=1}^{d^{min}}p(\bm y_{\bm r j} \mid \bar{\bm y}_{\bm r j-1}, \bm r \neq \bm 1, d^{min} \neq d^{max}) \\
& \prod_{j=d^{min}+1}^{d^{max}}p(\bm y^{max}_{\bm r j} \mid \bar{\bm y}_{\bm r d^{min}},\bm y^{max}_{\bm r d^{min}+1}, \ldots, \bm y^{max}_{\bm r d^{max}-1}, \bm r \neq \bm 1, d^{min} \neq d^{max}),
\end{align*}
where $\bm y^{max}_{\bm r j}$ denote the observed outcomes at time $j$ associated with the latest dropout time in pattern $d^{min}\neq d^{max}$. These distributions are again identified under partial ignorability for $d^{min}$ and $d^{max}$. The observed data part is obtained by integrating out the intermittent missing outcomes~$\bm y_{\bar{\bm r}j}$ for $j \leq d^{min}$ and the intermittent missing outcomes $\bm y^{max}_{\bar{\bm r}j}$ for~$d^{min}< j \leq~d^{max}$. 
\end{itemize}
Two components form the the extrapolation distribution: 
\begin{itemize}
\item[1] The unidentified outcome distributions among the non-completers with $d^{min}=d^{max}$: 
\begin{equation*}
p(\bm y_{\bar{\bm r}} \mid \bm y_{\bm r}, \bm r \neq \bm 1, d^{min}=d^{max}) = \prod_{j=d^{max}+1}^{J}p(\bm y_{\bar{\bm r} j} \mid \bar{\bm y}_{\bm r j-1}, \bm r \neq \bm 1, d^{min}=d^{max}),
\end{equation*}
which are unidentified for both utilities and costs at $j>d^{max}$. 

\item[2] The unidentified outcome distributions among the non-completers with $d^{min} \neq d^{max}$:
\begin{align*}
p(\bm y_{\bar{\bm r}} \mid \bm y_{\bm r}, \bm r \neq \bm 1, d^{min} \neq d^{max}) = & \prod_{j=d^{max}+1}^{J}p(\bm y_{\bar{\bm r} j} \mid \bar{\bm y}_{\bm r j-1}, \bm r \neq \bm 1, d^{min} \neq d^{max}) \\
& \prod_{j=d^{min}+1}^{d^{max}}p(\bm y^{min}_{\bar{\bm r} j} \mid \bar{\bm y}_{\bm r d^{min}}, \bm y^{max}_{\bm r d^{min}+1},\ldots, \bm y^{max}_{\bm r d^{max}}, \bm r \neq \bm 1, d^{min} \neq d^{max}),
\end{align*}
where $\bm y^{min}_{\bm r j}$ denote the observed outcomes at time $j$ associated with the earliest dropout time in pattern $d^{min}\neq d^{max}$. These distributions are unidentified for both types of outcomes at $j>d^{max}$ and only for the outcome associated with the earliest dropout time $\bm y^{min}_{\bm r j}$ at $d^{min}< j \leq d^{max}$ in pattern $d^{min}\neq d^{max}$.
\end{itemize}
More generally, the extrapolation factorisation for $p(\bm y, \bm r)$ conditional on $\bm d=(d^{min},d^{max})$~is~given~by:
\begin{align*}
\begin{split}
p(\bm y, \bm r \mid \bm \omega)&=p(\bm r \mid  \bm \pi)\left[p(\bm y \mid \bm r=\bm 1, \bm \lambda)\right]^{\mathbb{I}(d^{min}=d^{max}=J)}\\
& \left[ \prod_{\;\;\;\;d^{min}=d^{max}}p(\bm y_{\bm r} \mid \bm r \neq \bm 1, \bm d,\bm \eta)\right]^{\mathbb{I}(d^{min}=d^{max}<J)}\\
& \left[ \prod_{\;\;\;\;d^{min}\neq d^{max}}p(\bm y_{\bm r} \mid \bm r \neq \bm 1, \bm d,\bm \upsilon)\right]^{\mathbb{I}(d^{min}\neq d^{max})}
\end{split}  \;\;\;\;\;\;\;\;\;\;\; \vast \} \;\;\;\;\;  \text{observed data distribution} \\
\begin{split}
    {}& \left[ \prod_{\;\;\;\;d^{min}=d^{max}}p(\bm y_{\bar{\bm r}} \mid \bm y_{\bm r}, \bm r \neq \bm 1, \bm d,\bm \xi)\right]^{\mathbb{I}(d^{min}=d^{max}<J)}\\
       & \left[ \prod_{\;\;\;\;d^{min}\neq d^{max}}p(\bm y_{\bar{\bm r}} \mid \bm  y_{\bm r},  \bm r \neq \bm 1, \bm d,\bm \zeta)\right]^{\mathbb{I}(d^{min}\neq d^{max})}
\end{split}  \;\;\;\;\;\; \vastt \} \;\;\;\;\;  \text{extrapolation distribution} 
\end{align*}
where all these components were defined above. 

The parameters indexing the full model $\bm \omega=(\bm \theta, \bm \pi)$ are defined such that: 1) $\bm \lambda$, $\bm \eta$ and $\bm \upsilon$ respectively denote the subsets of $\bm \theta$ that index the observed data among the completers, those who drop out at the same time for both outcomes and those who have different dropout times for $\bm u$ and $\bm c$; 2) $\bm \xi$ and $\bm \zeta$ respectively denote the subsets of $\bm \eta$ and $\bm \upsilon$ that index the missing data in the corresponding patterns. The joint distribution has five components. The first is given by the model for the patterns and the model for the completers ($\bm r=\bm 1$), where no missingness occurs. The second and third components are the models for the observed data $\bm y_{\bm r}$ across $\bm r \neq \bm 1$ where $d^{min}=d^{max}$ and $d^{min}\neq d^{max}$ which, together with the first component, form the observed data distribution. The last two components are the models for the missing data $\bm y_{\bar{\bm r}}$ across $\bm r \neq \bm 1$ where $d^{min}=d^{max}$ and $d^{min}\neq d^{max}$, which form the extrapolation distribution. We will exploit this representation of the model in Section~\ref{mis_response} to define our identification strategy of the extrapolation distribution using different combinations of identifying restrictions and sensitivity parameters.

\subsection{Model for Dropout and the Observed Outcome}\label{obs_response}
The distribution of the dropout indicators $\bm D_i = (D^{min}_{i}, D^{max}_{i})$ is modelled using a multinomial distribution defined on~$\{1,\ldots,J^2 \}$, where $J^2$ is the total number of dropout patterns in the study. We note that, although the dropout patterns can be many in theory, this number is typically limited by the fact that most health technology assessment studies have few follow-up points. The dropout probabilities $\bm \pi_{t_{i}}=P(D^{min}_{i}=d^{min}, D^{max}_i = d^{max})$ depend on the treatment indicator $t_i$, where $t_i=1$ for the control and $t_i=2$ for the intervention group. We specify a prior for $\bm \pi_{t_{i}}$ that incorporates some knowledge about the expected dropout probabilities for the outcomes.
First, we give more weight to the completers ($\bm D_i=\bm J$) compared to all dropout patterns to reflect the idea that, given an expected dropout rate of $x\%$, we expect at least $(1-x)\%$ of the individuals to provide complete data. Second, we believe it more likely that individuals who leave the study will drop out at the same time for both outcomes ($D^{min}_i=D^{max}_i$) with respect to dropping out at different times for the two types of outcomes ($D^{min}_i<D^{max}_i$). We incorporate this prior information on $\bm \pi_{t_{i}}$ using the following Dirichlet distribution:
\begin{equation}\label{prior_dir}
\bm \pi_{t_{i}} \sim  \text{Dirichlet}\left(1-x,\frac{x/2}{J-1}, \ldots, \frac{x/2}{J-1},\frac{x/2}{(J^2 - J)},\ldots,\frac{x/2}{(J^2 - J)}\right),
\end{equation}
where the different prior probabilities, typically referred to as \textit{concentration} parameters, have the following interpretation: $1-x$ is the prior probability for completing the study; $\frac{x/2}{J-1}$ is the prior dropout probability assigned to each of the $J-1$ patterns where $D^{min}_i=D^{max}_i$; $\frac{x/2}{(J^2- J)}$ is the prior dropout probability assigned to each of the remaining $J^2-J$ patterns where $D^{min}_i\neq D^{max}_i$. 

We increase the weight of the prior on posterior inferences by multiplying each concentration parameter by a constant $Q>1$. This parameterisation assumes that samples will centre around the prior probabilities for each type of dropout patterns. We note that an alternative prior specification to (\ref{prior_dir}) would be to use a multinomial regression to model the conditional dependence between the two dropout indicators. However, interpretation of the parameters of this regression is difficult and model convergence is typically slow. For comparison, we also consider a prior for $\bm \pi_{t_{i}}$ based on a noninformative $\text{Dirichlet}(1,\ldots,1)$. For each part of the model, we assume different sets of parameters for the treatments, but to simplify notation we suppress the dependence on $t_i$ in the~following.

A critical problem in the statistical analysis of health technology assessment data is the choice of the model for $\bm y=(\bm u, \bm c)$, given that both utilities and costs typically present some complexities (e.g.~correlation and non-normality) that should be addressed using appropriate methods. Different approaches have been proposed in the literature to deal with these statistical issues and, particularly within a Bayesian approach, the use of more appropriate parametric modelling has been recommended to improve the model fit, either for costs~\citep{OHagan,Ng}, utilities~\citep{Basu,Gabriob}, or both~\citep{baio2014bayesian,gabrio2020bayesian}. Given the complexity of characterising the joint distribution of cost-effectiveness data with a parametric approach, we recommend the use of a nonparametric approach. Thus, we model the observed outcomes $\bm y_{ij}=(u_{ij},c_{ij})$ through a Dirichlet process mixture of normals~\citep{Escobar1995}. The model can be represented as
\begin{align}\label{DP_rep}
\bm y_i &\sim \text{Normal}\left(\bm \mu_{i}, \bm \Sigma_{i} \right), \nonumber \\
\left(\bm \mu_{i}, \bm \Sigma_{i} \right) & \sim G,\\
G & \sim DP(\alpha,G_0). \nonumber
\end{align} 
The set of complete outcomes is assumed to come from a multivariate normal distribution, indexed by the mean vector $\bm \mu_{i}$ and covariance matrix $\bm \Sigma_{i}$, which follow a Dirichlet process $G$, indexed by the baseline distribution $G_0$ and the concentration parameter~$\alpha$. Dependence is introduced using the stick-breaking representation of each element in this set~\citep{sethuraman1994constructive}, which allows us to re-write and approximate the model as a finite mixture of normals~\citep{ishwaran2002approximate}:
\begin{equation}\label{approx_dp}
\bm y_i \sim \sum_{k=1}^K\nu^{(k)} \text{Normal}\left(\bm \mu^{(k)},\bm \Sigma^{(k)} \right),
\end{equation}
where $\nu^{(k)}=V^{(k)}\prod_{j<k}\left(1-V^{(j)} \right)$, with $V^{(k)} \sim \text{Beta}(1,\alpha)$, for $k=1,\ldots,K$. When $K=1$, the model corresponds to a multivariate normal distribution while for $K$ large ($\infty$) the approximation to the Dirichlet process is~improved. 

A convenient reparameterisation of the model above can be achieved through the \textit{generalised autoregressive coefficients} (GARP) and \textit{innovation variances} (IV) decomposition of the multivariate distribution within each mixture component $k$~\citep{Taddy2008, Linero2015}. In particular, for each mixture component $k$, we implement the GARP/IV parameterisation to express the joint normal distribution of $\bm y_i$ as the product of a sequence of normal distributions. To simplify the specification of the model we assume a first-order Markov dependence structure for the data and obtain the factorisation
\begin{equation}\label{model_y_fact}
p(\bm y_{i})=f^{(k)}(\bm y_i)=f^{(k)}(\bm y_{i1})\prod_{j=2}^J f^{(k)}(\bm y_{ij} \mid \bm y_{ij-1}),
\end{equation}
for $k=1,\ldots,K$. We note that, with the above reparameterisation, we are not restricting the complete data model to be first-order. Within each mixture component $k$, the distribution $f^{(k)}(\bm y_{ij} \mid \bm y_{ij-1})$ is the density of the bivariate response $\bm y_{ij}$ at time $j$ for subjects who have their outcomes observed at time $j-1$ conditional on $\bm y_{ij-1}$. We note that, compared with standard practice in trial-based economic evaluations, where missing values are either ignored or imputed at an aggregated level, the proposed model represents a considerable improvement which captures the longitudinal aspect as well as potential higher order dependence relationships in the data. 

The baseline distributions $f^{(k)}(\bm y_{i1})$ from (\ref{model_y_fact}) are specified as
\begin{equation}\label{base_dist_y}
f^{(k)}(\bm y_{i1}) \sim \mbox{Normal}\left(\bm \gamma^{(k)}_{1}, \bm \Sigma^{(k)}_{1}\right),
\end{equation}
where $\bm \gamma^{(k)}_{1}$ is a bivariate parameter associated with the utility and cost baseline means, while $\bm \Sigma^{(k)}_{1}$ is the corresponding $2 \times 2$ covariance matrix. The identified distributions $f^{(k)}(\bm y_{ij} \mid \bm y_{ij-1})$ at time $j>1$ are specified as
\begin{equation}\label{followup_dist_y}
f^{(k)}(\bm y_{ij} \mid \bm y_{ij-1}) \sim \mbox{Normal}\left(\bm \gamma^{(k)}_{j} + \bm \phi^{(k)}_{j}\left(\bm y_{ij-1} - \bm \gamma^{(k)}_{j-1}\right), \bm \Sigma^{(k)}_{j}\right),
\end{equation}
where $\bm \gamma^{(k)}_{j}$ is the bivariate mean parameter at $j$, $\bm \phi^{(k)}_{j}$ are bivariate regression coefficients controlling the dependence between responses at $j$ and $j-1$, and $\bm \Sigma^{(k)}_{j}$ is the $2 \times 2$ covariance matrix at $j$. Detailed information about prior specification for all model parameters is provided in the web appendix.
 
\subsection{Identification of the Extrapolation Distribution}\label{mis_response}
We identify the extrapolation distribution $p(\bm y_{\bar{\bm r}} \mid \bm y_{\bm r}, d(\bm r))$ using combinations of identifying restrictions and sensitivity parameters~\citep{little1994class,Daniels2008}. Identifying restrictions are modelling choices which are used to link the observed data distribution to the extrapolation distribution under a benchmark scenario, from which interpretable deviations are then explored via sensitivity parameters to assess how inferences are driven by our assumptions~\citep{Linero2018}. We define our benchmark scenario using different identifying restrictions for each component of the extrapolation distribution conditional on $\bm d =(d^{min},d^{max})$. 

Let $f_{j}(\bar{\bm y}_{j-1})$ denote the set bivariate outcome distributions up to $j-1$ for those individuals who drop out at time $j$ and let $f_{\geq j}(\bm y_{\bm rj} \mid \bar{\bm y}_{j-1})$ denote the set of observed outcome distributions at $j$ for those who drop out at $j,j+1,\ldots,J$ for $j\geq 2$. Then, using the MAR or \textit{available case missing value} restriction~\citep{little1995modeling}, the unidentified distributions at time $j>d^{max}$ for the patterns where $d^{min}=d^{max}=d$ can be identified as: 
\begin{equation}\label{acmv_p1}
f_{d}(\bm y_{\bar{\bm r} j} \mid \bar{\bm y}_{\bm r j-1}, d^{min}=d^{max}) = f_{\geq j}(\bm y_{\bm r j}\mid \bar{\bm y}_{\bm r j-1}) = \sum_{s=j}^J \frac{\pi_s f_s(\bar{\bm y}_{\bm r j-1})}{\sum_{l=j}^J \pi_{l} f_l(\bar{\bm y}_{\bm r j-1})} f_s(\bm y_{\bm r j} \mid \bar{\bm y}_{\bm r j-1}).
\end{equation}
Essentially, the unidentified distributions at $j > d^{max}$ are identified using a mixture over the distributions at the same time for all the identified patterns $s=j,\ldots,J$. The terms forming the ratio are the mixing coefficients, corresponding to $P(D=s \mid \bar{\bm y}_{\bm r j-1})$, i.e.~the probability of droput at $D=s$ given the observed outcome history up to $j-1$, for $s=j,\ldots,J$.

For the patterns $d^{min} \neq d^{max}$, two different dropout indicators ($d^{u},d^{c}$) exist according to which type of outcome the individuals drop out earlier ($d^{min}$) and later ($d^{max}$) from the study. Thus, for these patterns, two different identification strategies are implemented to identify: 1) the unidentified distributions for each type of outcome at time $j$ between the two dropout times within each pattern, for $d^{min} < j \leq d^{max}$ and $j\geq 2$; 2) the unidentified distributions for each type of outcome at time $j$ after the latest dropout time, for $j>d^{max}$ and $j \geq 2$. To distinguish the identified and unidentified components of the outcome distributions for the patterns $d^{min}\neq d^{max}$, we respectively denote with $\bm y^{min}_{\bm rj}$ and $\bm y^{min}_{\bar{\bm r}j}$, and with $\bm y^{max}_{\bm rj}$ and $\bm y^{max}_{\bar{\bm r}j}$, the observed and missing components at time $j$ for the outcome associated with the earliest and latest dropout, where $\bm y^{min}_j=(\bm y^{min}_{\bm rj},\bm y^{min}_{\bar{\bm r}j})$ and $\bm y^{max}_j=(\bm y^{max}_{\bm rj},\bm y^{max}_{\bar{\bm r}j})$. 

The unidentified distributions at time $d^{min}<j \leq d^{max}$ are identified using all the observed distributions at $j$ up to $d^{max}$, that is:
\begin{equation}\label{acmv_p2}
f_{\substack{d^{min}\\d^{max}}}(\bm y^{min}_{\bar{\bm r} j} \mid \bar{\bm y}^{min}_{\bm r j-1}, \bar{\bm y}^{max}_{\bm r d^{max}}, d^{min}\neq d^{max}) = f^{d^{max}}_{\geq j}(\bm y^{min}_{\bm r j} \mid \bar{\bm y}^{min}_{\bm r j-1}, \bar{\bm y}^{max}_{\bm r d^{max}})
\end{equation}
where the notation $f_{\substack{d^{min}\\d^{max}}}(\cdot)$ denotes the outcome distributions for the patterns associated with the two different dropout times $d^{min}$ and $d^{max}$, while the term $f^{d^{max}}_{\geq j}(\bm y^{min}_{\bm rj} \mid \bar{\bm y}^{min}_{\bm r j-1}, \bar{\bm y}^{max}_{\bm r d^{max}})$ denotes the set of observed outcome distributions $\bm y^{min}$ from time $j$ up to $d^{max}$. Thus, the extrapolation distribution is identified using a mixture over the observed distributions at the same time for all the identified patterns up to $d^{max}$, with $d^{min}\neq d^{max}$. 


Finally, for the patterns $d^{min} \neq d^{max}$, the unidentified distributions at time $j > d^{max}$ are~identified~as:
\begin{equation}\label{acmv_p3}
f_{\substack{d^{min}\\d^{max}}}(\bm y_{\bar{\bm r}j} \mid \bar{\bm y}_{\bm r j-1}, d^{min}\neq d^{max}) = f_{\geq j}(\bm y_{\bm r j} \mid \bar{\bm y}^{min}_{\bm r d^{min}}, \bar{\bm y}^{max}_{\bm r d^{max}},\ldots, \bm y_{\bm r j-1}),
\end{equation}
that is using a mixture over the observed distributions at the same time for all identified patterns after $d^{max}$, with $d^{min}\neq d^{max}$ and $j>d^{max}$. We note that (\ref{acmv_p3}) cannot be completely identified unless the unidentified distributions at $d^{min}<j \leq d^{max}$ are first identified using, for example, the restrictions in~(\ref{acmv_p2}).
We refer to the web appendix for a detailed presentation of the identifying restrictions used to identify the extrapolation distribution for the patterns $d^{min}=d^{max}$ and $d^{min}\neq d^{max}$.

\section{Analysis under Nonignorability}\label{mnar}

\subsection{Sensitivity Parameters}\label{sp}
To this point, we have identified the extrapolation distribution $p(\bm y_{\bar{\bm r}} \mid \bm y_{\bm r}, d(\bm r))$ under a benchmark scenario through a set of identifying restrictions as described in Section~\ref{mis_response}. However, although this approach represents a reasonable starting point, the assumption that missingness can be fully explained based on the observed data alone may be questionable. This is the case if, for example, we expect patients who leave the study to have worse health states (i.e.~lower utilities) than those that continue on, even after conditioning on observed measurements. Hence, we allow for departures from the benchmark assumption after dropout via sensitivity parameters for each type of outcome $\bm \Delta_{j}$~\citep{Wang2011,Gaskins2016}. We identify the post-dropout distributions by applying location shifts to the distributions identified under the benchmark scenario (Section~\ref{mis_response}) separately for the patterns $d^{min}=d^{max}$ ($\bm \Delta^{d^{min}=d^{max}}_j$) and the patterns $d^{min}\neq d^{max}$ ($\bm \Delta^{d^{min}\neq d^{max}}_j$). 
To that end, for the patterns $d^{min}=d^{max}$, we re-write (\ref{acmv_p1}) as $\sum_{s=j}^{J} \phi_{js}(\bar{\bm y}_{\bm r j-1})f_{s}(\bm y_{\bm rj}\mid \bar{\bm y}_{\bm rj-1})$, where $\phi_{js}(\bar{\bm y}_{\bm r j-1})$ denote the mixing coefficients $P(D=s\mid \bar{\bm y}_{\bm rj-1})$. Then, we identify the distributions at $j>d^{max}$ as:
\begin{equation}\label{model_u_mis1}
f_{d}(\bm y_{j} \mid \bm y_{j-1}, d^{min}=d^{max}) \equiv  \sum_{s=j}^{J} \phi_{js}(\bar{\bm y}_{\bm r j-1}) \tilde{f}_{s}(\bm y_{j} \mid \bm y_{j-1}, d^{min}=d^{max}),
\end{equation}
where $\tilde{f}_{s}(\bm y_{j} \mid \bm y_{j-1}, d^{min}=d^{max})$ incorporates the location shifts $\bm \Delta_{j}^{d^{min}=d^{max}}$ within the identified distributions and has mean $\mbox{E}[\bm Y_{j}\mid \bm Y_{j-1}, d^{min}=d^{max}] + \bm \Delta_{j}^{d^{min}=d^{max}}$. We note that the choice $\bm \Delta_{j}^{d^{min}=d^{max}}=\bm 0$ produces $f_{d}(\bm y_{j} \mid \bm y_{j-1}, d^{min}=d^{max})$ as the distributions under the benchmark scenario in~(\ref{acmv_p1}).

Similarly, for the patterns $d^{min}\neq d^{max}$, we re-write (\ref{acmv_p2}) as $\sum_{\substack{s^{min}=j\\s^{max}=d^{max}}}^{d^{max}} \phi_{j(\substack{s^{min}\\s^{max}})}(\bar{\bm y}^{min}_{\bm r j-1},\bar{\bm y}^{max}_{\bm r d^{max}})\allowbreak f_{\substack{s^{min}\\s^{max}}}(\bm y^{min}_{\bm rj}\mid \bar{\bm y}^{min}_{\bm rj-1},\bar{\bm y}^{max}_{\bm r d^{max}})$, where $\phi_{j(\substack{s^{min}\\s^{max}})}(\bar{\bm y}^{min}_{\bm r j-1},\bar{\bm y}^{max}_{\bm r d^{max}})$ denote the mixing coefficients $P(D^{min}=s^{min},D^{max}=s^{max}\mid \bar{\bm y}^{min}_{\bm rj-1},\bar{\bm y}^{max}_{\bm r d^{max}})$. Then, we identify the distributions at $d^{min}<j \leq d^{max}$ as:
\begin{equation}\label{model_u_mis2}
f_{\substack{d^{min}\\d^{max}}}(\bm y^{min}_{j} \mid \bm y_{j-1}, d^{min} \neq d^{max}) \equiv \sum^{d^{max}}_{\substack{s^{min}=j\\s^{max}=d^{max}}} \phi_{j(\substack{s^{min}\\s^{max}})}(\bar{\bm y}^{min}_{\bm r j-1},\bar{\bm y}^{max}_{\bm r d^{max}}) \tilde{f}_{\substack{s^{min}\\s^{max}}}(\bm y^{min}_{j} \mid \bm y_{j-1}, d^{min} \neq d^{max}),
\end{equation}
where $\tilde{f}_{\substack{s^{min}\\s^{max}}}(\bm y^{min}_{j} \mid \bm y_{j-1}, d^{min} \neq d^{max})$ incorporates the  location shifts $\bm \Delta_{j}^{d^{min}\neq d^{max}}$ within the identified distributions and has mean $\mbox{E}[\bm Y^{min}_{j}\mid \bm Y_{j-1}, d^{min}\neq d^{max}] + \bm \Delta_{j}^{d^{min}\neq d^{max}}$. Again, setting $\bm \Delta_{j}^{d^{min}\neq d^{max}}=\bm 0$ produces $f_{\substack{d^{min}\\d^{max}}}(\bm y^{min}_{j} \mid \bm y_{j-1}, d^{min}\neq d^{max})$ as under the benchmark scenario in (\ref{acmv_p2}).

Finally, we also re-write (\ref{acmv_p3}) using a similar identification strategy as in (\ref{model_u_mis1}), but including location shifts $\bm \Delta_{j}^{d^{min}\neq d^{max}}$ specific to the patterns $d^{min}\neq d^{max}$:
\begin{equation}\label{model_u_mis3}
f_{\substack{d^{min}\\d^{max}}}(\bm y_{j} \mid  \bm y_{j-1}, d^{min}\neq d^{max}) \equiv  \sum_{\substack{s^{min}=j\\s^{max}=j}}^{J} \phi_{j(\substack{s^{min}\\s^{max}})}(\bar{\bm y}_{\bm r j-1}) \tilde{f}_{\substack{s^{min}\\s^{max}}}(\bm y_{j} \mid \bm y_{j-1}, d^{min}\neq d^{max}),
\end{equation}
where $\tilde{f}_{\substack{s^{min}\\s^{max}}}(\bm y_{j} \mid \bm y_{j-1}, d^{min}\neq d^{max})$ has mean $\mbox{E}[\bm Y_{j}\mid \bm Y_{j-1}, d^{min} \neq d^{max}] + \bm \Delta_{j}^{d^{min}\neq d^{max}}$. However, in contrast to (\ref{model_u_mis1}), the post-dropout distributions for the patterns $d^{min}\neq d^{max}$ at time $j>d^{max}$ are identified only after the unidentified distributions at $d^{min}<j \leq d^{max}$ are first identified using~(\ref{model_u_mis2}). We note that different sets of sensitivity parameters can be potentially used to identify the extrapolation distribution conditional on the dropout patterns and time of dropout.

\subsection{Computation and Inference}\label{computation}
Recall that our modelling strategy consists of a working model $p^\star(\bm y, d(\bm r) \mid \bm \omega)$ for the joint distribution of the outcome and missingness patterns. The model $p(\bm r \mid \bm \pi)$ is specified using a Multinomial distribution, while the model $p(\bm y \mid \bm r, \bm \theta)$ is specified separately by dropout pattern using an approximation of the Dirichlet process mixture upon truncating the stick-breaking construction at a finite number of components $K$, with $\bm \omega = (\bm \pi,\bm \theta)$. Posterior computation is carried out at each iteration of the posterior distribution. 

First, we draw samples $(\bm \pi, \bm \theta)$ from the posterior distribution given the observed data $(\bm y_{\bm r}, d(\bm r))$. When the working model framework in Section~\ref{framework} is used, samples of $(\bm \pi, \bm \theta)$ can be obtained by fitting the working model $p^{\star}$ by data augmentation, taking advantage of the fact that $p(\bm y_r, d(\bm r) \mid \bm \omega)=\int p(\bm y, d(\bm r)\mid \bm \omega)d \bm y_{\bar{\bm r}}$. Second, inference is done by computing the effects of interest directly from the posterior samples and the chosen identifying restriction. Specifically, once the samples of $\bm \omega=(\bm \pi, \bm \theta)$ are obtained, we are interested in the computation of desired functionals of the form:
\begin{equation}\label{func_comp}
\mbox{E}\left[ \mathcal{T}(\bm Y) \mid \bm \omega \right]=\int \mathcal{T}(\bm y)p(\bm y \mid \bm\omega)d\bm y,
\end{equation}
which are typically not available in closed form as they depend on $p(d(\bm r) \mid \bm \omega)=p(d^{min},d^{max}\mid \bm \omega)$ and $\mbox{E}\left[ \mathcal{T}(\bm Y) \mid d^{min}, d^{max}, \bm \omega \right]$. Indeed, the specific form of the expectation $\mbox{E}\left[ \mathcal{T}(\bm Y) \mid d^{min}, d^{max}, \bm \omega \right]$ depends on our assumption about the missing data. For example, for the patterns $d^{min}=d^{max}=d$ under the MAR restrictions, we have:
\begin{equation*}\label{func_comp_p1}
\begin{split}
\footnotesize
\mbox{E}\left[\mathcal{T}(\bm Y) \mid d^{min}=d^{max}, \bm \omega \right] = & \int \mathcal{T}(\bm y) p_{d}(\bar{\bm y}_{d} \mid \bm \omega)  p_{\geq d+1}(\bm y_{d+1}\mid \bar{\bm y}_{d}, \bm \omega) p_{\geq d+2}(\bm y_{d+2}\mid \bar{\bm y}_{d+1}, \bm \omega) \cdots  \\
& \cdots p_{J}(\bm y_{J}\mid \bar{\bm y}_{J-1}, \bm \omega) d \bm y,
\end{split}
\end{equation*}
while for the patterns $d^{min}\neq d^{max}$, under the restrictions described in Section~\ref{mis_response}, we have:
\begin{equation*}\label{func_comp_p2}
\begin{split}
\footnotesize
\mbox{E}\left[\mathcal{T}(\bm Y) \mid d^{min}\neq d^{max},\bm \omega \right] = & \int \mathcal{T}(\bm y) p_{d^{min}}(\bar{\bm y}_{d^{min}} \mid \bm \omega) \\
& p_{d^{min}+1}(\bm y^{max}_{d^{min}+1} \mid \bar{\bm y}_{d^{min}}, \bm \omega) p_{d^{min}+2}(\bm y^{max}_{d^{min}+2} \mid \bar{\bm y}_{d^{min}}, \bm y^{max}_{d^{min}+1}, \bm \omega) \cdots \\
& p_{d^{max}}(\bm y^{max}_{d^{max}} \mid \bar{\bm y}_{d^{min}}, \bm y^{max}_{d^{min}+1}, \ldots, \bm y^{max}_{d^{max}-1}, \bm \omega) \\
& p_{\geq d^{min}+1}^{d^{max}}(\bm y^{min}_{d^{min}+1} \mid \bar{\bm y}_{d^{min}}, \bm y^{max}_{d^{min}+1},\ldots, \bm y^{max}_{d^{max}}, \bm \omega) \\
& p_{\geq d^{min}+2}^{d^{max}}(\bm y^{min}_{d^{min}+2} \mid \bar{\bm y}_{d^{min}+1}, \bm y^{max}_{d^{min}+2},\ldots, \bm y^{max}_{d^{max}}, \bm \omega) \cdots \\
& \cdots p_{\geq d^{max}}^{d^{max}}(\bm y^{min}_{d^{max}} \mid \bar{\bm y}_{d^{max}-1}, \bm y^{max}_{d^{max}}, \bm \omega) \\
& p_{\geq d^{max}+1}(\bm y_{d^{max}+1}\mid \bar{\bm y}_{d^{max}}, \bm \omega) p_{\geq d^{max}+2}(\bm y_{d^{max}+2}\mid \bar{\bm y}_{d^{max}+1}, \bm \omega)  \cdots \\ 
& \cdots p_{J}(\bm y_{J}\mid \bar{\bm y}_{J-1}, \bm \omega) d \bm y.
\end{split}
\end{equation*}
We calculate $\mbox{E}\left[\mathcal{T}(\bm Y) \mid d^{min}=d^{max}, \bm \omega \right]$ and $\mbox{E}\left[ \mathcal{T}(\bm Y) \mid d^{min}\neq d^{max},\bm \omega \right]$ using Monte Carlo integration by sampling pseudodata $\bm y^{\star}=(\bm y^{(1)}, \ldots, \bm y^{(L)})$ and $d(\bm r)^{\star}=(d(\bm r)^{(1)},\ldots,d(\bm r)^{(L)})$, and forming the average $L^{-1}\sum_{l=1}^L \mathcal{T}(\bm y^{(l)})$, separately for the patterns $d^{min}=d^{max}$ and $d^{min}\neq d^{max}$, for some large~$L$. We note that this is essentially an application of G-computation within the Bayesian paradigm~\citep{scharfstein2014global,Linero2015}. Detail of the G-computation procedure used to calculate the desired effects is provided in the web appendix.

\subsection{Priors on the Sensitivity Parameters}\label{priors_sp}
Although in Section~\ref{mnar} we specify sensitivity parameters $\bm \Delta^{d^{min}=d^{max}}_j$ and $\bm \Delta^{d^{min}\neq d^{max}}_j$ conditional on the dropout patterns, to ease notation, throughout this section we use the term $\bm \Delta_j$ to denote any sensitivity parameter included in our nonignorbale analysis. The same prior specification described in this section is then applied separately for each pattern-specific sensitivity parameters to specify informative departures from our benchmark scenario.

We specify informative priors on all sensitivity parameters $\bm \Delta_j \neq \bm 0$, which are calibrated using information on the scale of the data as an intuitive starting point~\citep{Daniels2008,Linero2015}. The residual standard deviation in the observed utilities and costs pooled across time and treatment groups is roughly $0.36$ and $\pounds 1500$ respectively, and it is thought unlikely that deviations from the benchmark would exceed a standard deviation for both types of responses. We note that, given the different amounts of dropout and the different proportions of dropout by treatment group, the same values of $\bm \Delta_j$ will affect each treatment differently. We marginally specify the prior distributions of each sensitivity parameter and then separately model their multivariate dependence structure through a \textit{copula model} to account for the possible time dependence between~$\bm \Delta_j$. Copulae allow to separate the modelling of the marginal distributions of each random variable from the modelling of their dependence structure~\citep{durante2013topological}. In general, it is possible to extract a copula from any joint distribution~$F(\cdot)$:
\begin{equation*}\label{copula}
C(u_1,\ldots,u_J) = F(F_1^{-1}(u_1),\ldots,F_J^{-1}(u_J)),
\end{equation*}
where $C$ denotes the copula, while $F(\cdot)$ and $F_j(\cdot)$ denote the joint and marginal cumulative distribution functions of the $J$ random variables $u_j$. When $F(\cdot)$ is chosen as $\bm \Phi$, the cdf of a multivariate normal distribution, then $C$ is a normal copula with mean vector $\bm \mu$ and covariance matrix $\bm \Sigma$. Since copulae are invariant to monotonic transformation of the margins, the dependence structure is entirely determined by the correlation matrix $\bm \Omega$, where $\bm \Sigma=\text{diag}(\bm \sigma) \times \bm \Omega \times \text{diag}(\bm \sigma)$. In our analysis, we model the dependence of $\bm \Delta_j$ through a normal copula, whose mean and correlation structures are defined according to a AR(1) or \textit{first-order autoregressive} structure. More specifically, the sensitivity parameter $\bm \Delta_j$ are linked at the mean level through a lag-one dependence process:
\begin{equation}\label{ar1_proc}
\bm \Delta_j \mid \bm \Delta_{j-1}  \sim \text{Normal}\left(\bm \Delta_0 + \bm \rho \bm \Delta_{j-1}, \bm \sigma_{\Delta_j}\right),
\end{equation}
in which the expected value of $\bm \Delta_j$ depends on an offset $\bm \Delta_0$ (a grand average expected shift after dropout) and on its value at $j-1$ through the linear expression in Equation~\ref{ar1_proc}. The parameter $\bm \rho=(\rho^u,\rho^c)$ is the two-vector of autoregressive coefficients capturing the temporal dependence between the shifts after dropout. The two-vector parameter $\bm \sigma_{\Delta_j}=(\sigma^u_{\Delta_j},\sigma^c_{\Delta_j})$ is the standard deviation of $\bm \Delta_j$, which is set to the empirical values $\text{sd}(\bm y)$ obtained from the corresponding observed utility and cost data across all times. 
We specify an AR(1) stationary process by imposing $\mid \bm \rho \mid <\bm 1$ and start the process at $\bm \Delta_{j}=\bm 0$, which corresponds to the time of dropout. Finally, the correlation matrix $\bm \Omega$ is parametrised by $\bm\rho_{\text{time}}$ and has the following~form:
\begin{equation*}
\bm \Omega = 
\begin{pmatrix}
1 & \bm\rho_1 & \cdots & \bm \rho_{\mid j-s \mid}\\
\bm\rho_1 & 1 & \cdots & \vdots \\
\vdots & \cdots & 1 & \vdots \\
 \bm\rho_{\mid j-s \mid} & \cdots & \cdots & 1
\end{pmatrix}
\end{equation*}
in which the exponent on the correlation declines linearly according to the time lag $(\mid j-s \mid)$. 

We explore alternative assumptions about the magnitude of the autocorrelation structure for $\bm \Delta_j$ by giving $\bm \rho$ point-mass priors and varying the priors along a discrete set. To ease the display of the inferences, we assume $\bm \rho$ to be the same across treatments and we vary the parameter over the grid $\{0.1,0.5,0.9\}$. We only consider positive values for $\bm \rho$ as we believe that, after dropout, individuals are associated with a progressive increase in their mean utility decrement/cost increment over time.


\section{Application}\label{results}

\subsection{Computation}
We fitted the model in \texttt{Stan}~\citep{carpenter2017stan} which is a software specifically designed for the analysis of Bayesian models using a type of \textit{Markov Chain Monte Carlo} (MCMC) algorithm known as \textit{Hamiltonian Monte Carlo}~\citep{Brooks}, and which is interfaced with \texttt{R} through the package \texttt{rstan}~\citep{team2016rstan}. We ran two chains with $25000$ iterations and a warm up of $5000$ iterations, for a total sample of $40000$ iterations for posterior inference. For selected key unknown quantities in the model, we assessed convergence and autocorrelation of the MCMC simulations using diagnostic measures including the \textit{potential scale reduction factor} and the \textit{effective sample size}~\citep{Gelman2}. The \texttt{Stan} code used to fit the model to the PBS data is provided in the web appendix. The entire \texttt{R} code is also available on the~\href{https://github.com/AnGabrio/Code/tree/master/dpm}{GitHub repository} of the authors.

\subsection{Model Assessment}\label{assessment}
We compute three relative measures of predictive accuracy to assess the fit of the proposed model with respect to alternative specifications. The first two are: the \textit{Widely Applicable Information Criterion} (WAIC)~\citep{watanabe2010asymptotic} and the \textit{Leave-One-Out Information Criterion} (LOOIC)~\citep{vehtari2017practical}. These information criteria are obtained based on the model deviance and a penalty for model complexity known as effective number of parameters ($p_{D}$). We also consider the \textit{Log-Pseudo Marginal likelihood} (LPML)~\citep{Geisser1979}, which is an estimator for the log marginal likelihood obtained by summing up the log of the Conditional Predictive Ordinate or CPO statistics~\citep{Chen2012}. Although not created as an information criterion, the LPML shares many of the same model selection uses as other information criteria and WAIC has been proven to be asymptotically equivalent to LPML~\citep{watanabe2010asymptotic}. To favour comparison across all measures, we compute a LPML criterion on the deviance scale, i.e. $\text{LPML IC}= -2 \times \text{LPML}$, so that the "best" model among those assessed is the one associated with smallest values.

We compare the proposed Dirichlet Process Mixture model (DPM) with three parametric specifications: 1) a multivariate Normal distribution fitted jointly to all variables (MVN); 2) the multivariate parametric model proposed by~\citet{gabrio2020bayesian}, using Beta distributions for the utility and Gamma distributions for the cost variables (BG); 3) a similar model to $2$ but replacing Gamma with LogNormal distributions for the costs (BLN). For each measure, the results between the models are obtained based on the observed data distribution as defined in Section~\ref{mis_response}. The estimated quantities are reported in~Table~\ref{tabpic}.
\begin{table}[H]
\centering
    \begin{tabular}{c|ccc} 
\toprule
      \textbf{Model} & \textbf{WAIC} & \textbf{LOOIC} & \textbf{LPML IC}\\
      \midrule
      MVN &  13513 &  13327  & 13326\\
      BG &  12696 &  12518  & 12520\\
      BLN &  11850 &  10918 &  10920\\
      DPM &  8154  &  8155 & 8154\\
\bottomrule
    \end{tabular}
    \caption{Values of three alternative measures of fit for each model assessed.}\label{tabpic}
\end{table}
All measures indicate that DPM is associated with a considerably better fit (lower values for all criteria) compared with all the other parametric models with MVN, not surprisingly, being associated with the worst performance. 

We also assess the absolute fit of the model using posterior predictive checks based on data replications, which are obtained by repeatedly sampling from the posterior predictive distribution~\citep{Xu2016}. First, at iteration $l$, we sample $\bm \omega^{(l)}$ from the data-augmented posterior $p(\bm \omega^{(l)} \mid \bm y_{\bm r}, \bm y^{(l-1)}_{\bar{\bm r}},d(\bm r))$, with $\bm y^{(l)}_{\bar{\bm r}}$ being drawn from $p(\bm y^{(l)}_{\bar{\bm r}} \mid \bm y_{\bm r}, d(\bm r), \bm \omega^{(l)})$. We note that, although the model is defined in terms of dropout indicators $d(\bm r)=(d^{min},d^{max})$, these are effectively functions of $\bm r$ so that, for each dropout pattern, the observed values of the missingness indicators $\bm r_j$ at each time $j$ can be retrieved. Next, we sample replicated data from $p(\tilde{\bm y}^{(l)},\tilde{d(\bm r)}^{(l)} \mid \bm \omega^{(l)})$ and keep the observed data replications $(\tilde{\bm y}^{(l)}, \tilde{d(\bm r)}^{(l)})$. These are defined as $\tilde{\bm y}_{\bm r}=\{ \tilde{y}^{min}_j : \tilde{d(r^{min}_j} = 1);   \tilde{y}^{max}_j : \tilde{d(r^{max}_j} = 1)\}$, where $(\tilde{y}^{min}_j,\tilde{y}^{max}_j)$ and $(\tilde{r}^{min}_j,\tilde{r}^{max}_j)$ are the components associated with the earliest and latest dropout outcome in pattern $d(\bm r)$. Thus, the replicated observed data are the components of $\tilde{\bm y}=(\tilde{\bm y}^{min},\tilde{\bm y}^{max})$ for which the corresponding replicated missingness indicators (based on the value of the replicated dropout indicators) are equal to~one.

\begin{figure}[H]
\centering
\subfloat[control]{\includegraphics[scale=0.45]{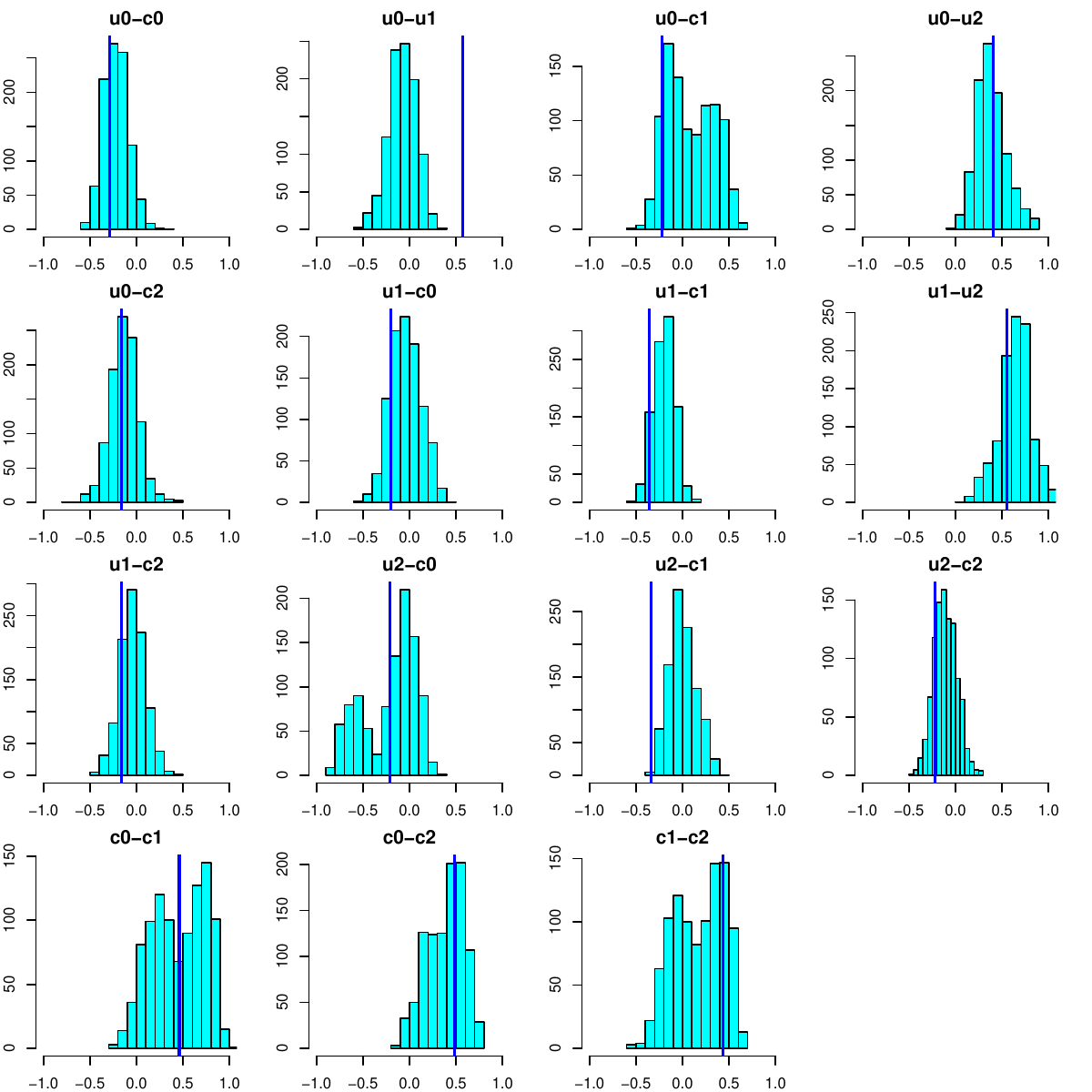}}\hspace*{0.75cm}
\subfloat[intervention]{\includegraphics[scale=0.45]{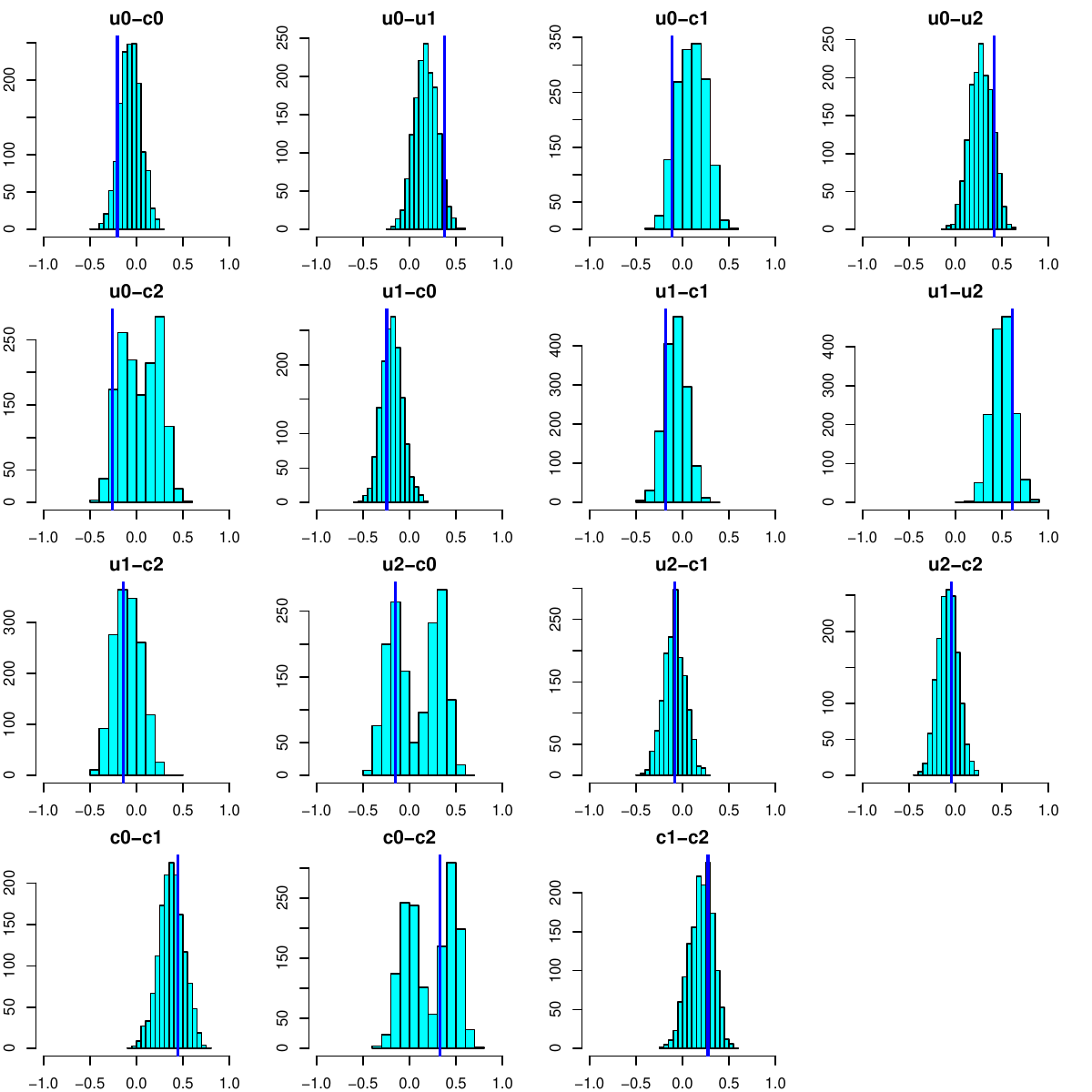}}
\caption{Posterior predictive distributions from DPM for the pairwise rank correlations between utilities and costs variables in the control (panel a) and intervention (panel b) arm for the replicated data (light blue bars) compared with the estimates based on the observed data in the real dataset (vertical blue~lines).}\label{ppc_corr}
\end{figure}

We compute the rank correlations between each pair of variables for each replicated dataset, and compare these with the corresponding values from the real dataset. The results, shown in Figure~\ref{ppc_corr}, suggest that the proposed model captures most of the correlations well both in the control (panel a) and intervention (panel b)~group. The only exception is represented by the replicated utility correlations between baseline and the first follow-up in the control group (panel a, second graph), which are systematically lower compared to the observed one.

\subsection{Results}\label{res}
This section summarises the results from the posterior distribution of the main quantities of interest, namely the marginal mean utility and cost parameters at each time point $(\mu^u_{jt}$, $\mu^c_{jt})$ and the marginal mean QALYs and total costs ($\mu_{et}$, $\mu_{ct}$) evaluated over the trial duration. Focus is given to the comparison of the results based on different missingness assumptions within the general modelling framework described in Section~\ref{framework}.

\begin{figure}[h!]
\centering
\subfloat[control]{\includegraphics[scale=0.52]{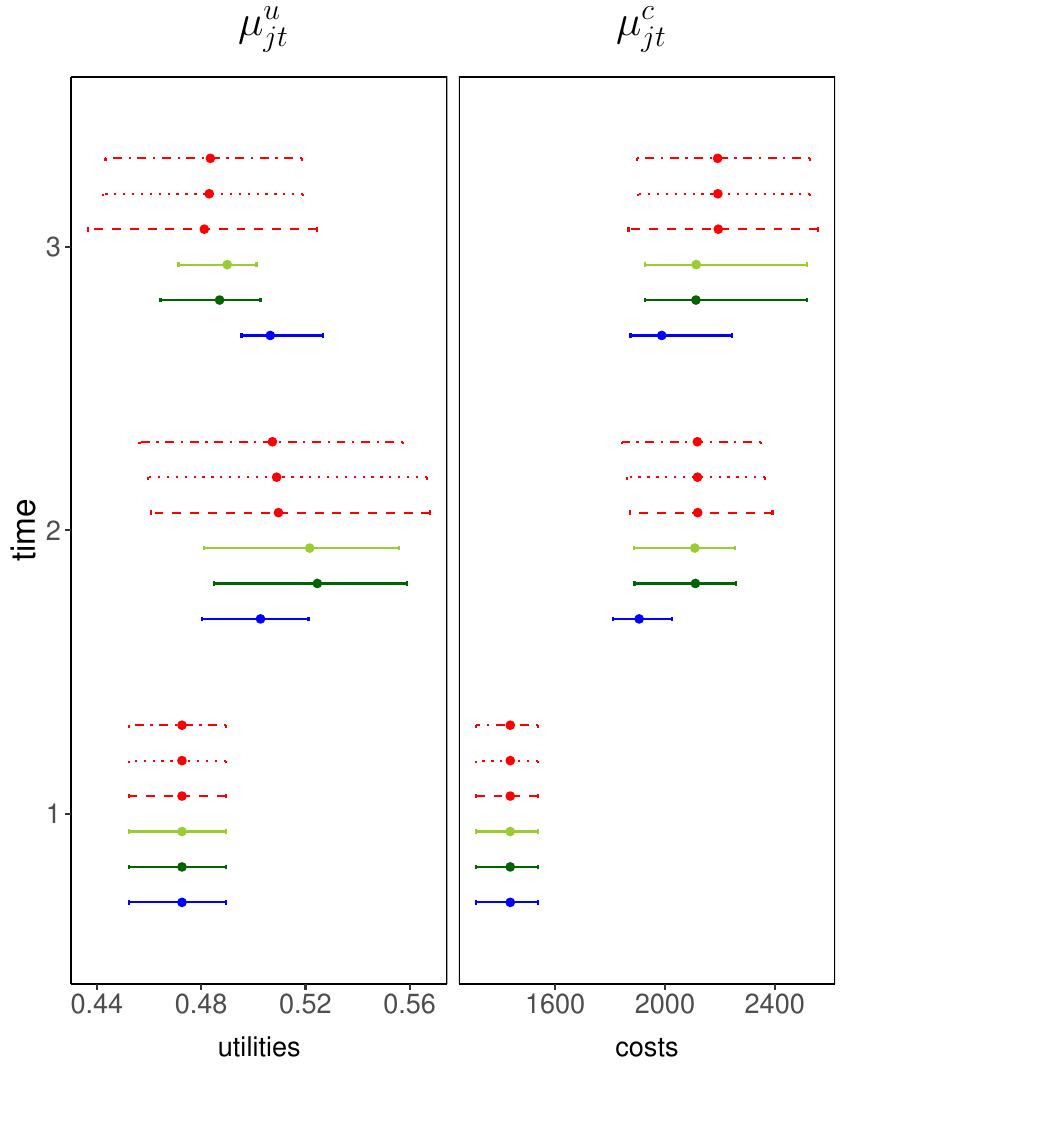}}
\subfloat[intervention]{\includegraphics[scale=0.52]{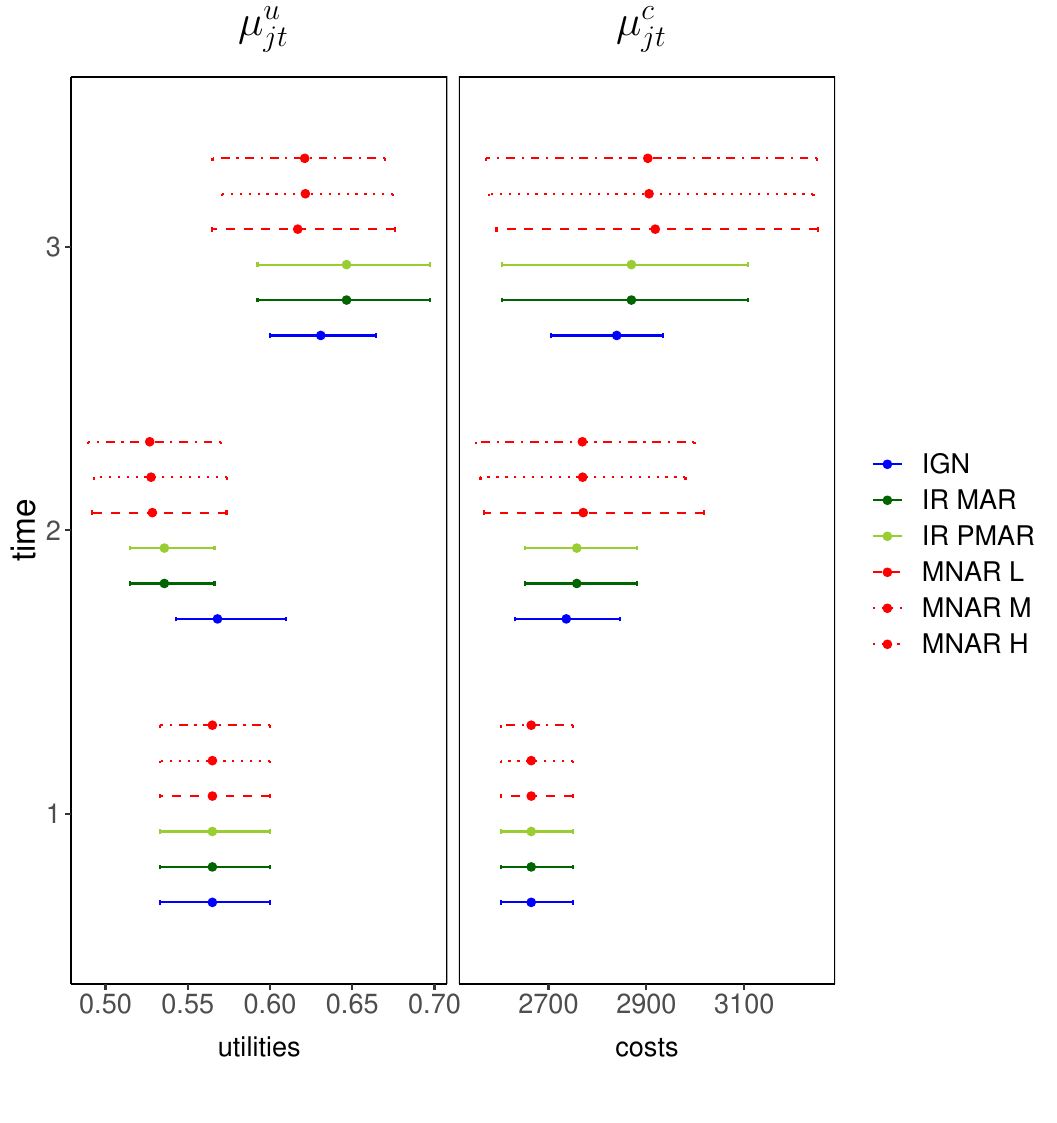}}
\caption{Posterior means and 95\% HPD intervals for the marginal utility and cost means in the control (panel a) and intervention (panel b) group at each time ($j=1,2,3$) in the study across alternative assumptions. Six scenarios are compared: nonignorable-MAR without restrictions (NOIG), ignorable based on MAR restrictions (IG), benchmark (BENCH), and MNAR scenarios based on a copula model with three alternative correlation values: $0.1$ (MNAR L), $0.5$ (MNAR M) and $0.9$ (MNAR~H).}\label{res_mean2}
\end{figure}

Figure~\ref{res_mean2} shows the posterior means and 95\% highest posterior density (HPD) credible intervals for the marginal utility and cost means $\bm \mu_{jt}=(\mu^u_{jt},\mu^c_{jt})$, obtained from fitting the model to all cases under six alternative scenarios: nonignorable MAR without using any identifying restrictions (NOIG), ignorable based on MAR identifying restrictions (IG), benchmark nonignorable based on identifying restrictions conditional on $d^{min}$ and $d^{max}$ (BENCH), and three nonignorable departure scenarios based on three different temporal correlation values for the sensitivity parameters: $0.1$ (MNAR L), $0.5$ (MNAR M) and $0.9$ (MNAR H). Results associated with each scenario are indicated with different colours and line types, separately reported for the control (panel a) and intervention (panel b)~groups.
The distributions of both $\bm \mu^u_j$ and $\bm \mu^c_j$ show values that are higher in the intervention compared with the control at each time $j$, and show relatively similar mean estimates across NOIG, IG and BENCH. However, with the exception of baseline ($j=0$), the utility/cost estimates under NOIG are consistently lower/higher compared to those under IG and BENCH. Across the three MNAR scenarios mean utilities and costs are respectively $4\%$ lower and $4.5\%$ higher on average compared to the estimates of the benchmark nonignorable scenario (BENCH), with relatively small variations between MNAR L, MNAR M and MNAR H. However, the HPD intervals for each marginal mean parameter do not show evidence of large discrepancies across all scenarios.

We then combined the cost and utility marginal means from the model and derived the corresponding QALYs and total costs marginal means $\bm \mu_{t}=(\mu_{et},\mu_{ct})$. Table~\ref{res_bar} shows the posterior means and 95\% HPD credible intervals associated with the targeted quantities under all~scenarios for both treatment~groups.
\begin{table}[H]
\caption{\label{res_bar}Posterior means and 95\% HPD credible intervals for $\mu_{et}$ and $\mu_{ct}$ in the control ($t=1$) and intervention ($t=2$) group under six alternative scenarios: NOIG, IG, BENCH and three MNAR scenarios (MNAR L, MNAR M, MNAR H).}
\centering
\scalebox{0.85}{
\begin{tabular}{c|cc|cc|cc|cc}
  \toprule
 \multirow{2}{*}{Scenario}   & \multicolumn{2}{c|}{$\mu_{e1}$} &  \multicolumn{2}{c|}{$\mu_{e2}$} &  \multicolumn{2}{c|}{$\mu_{c1}$} &  \multicolumn{2}{c}{$\mu_{c2}$} \\   \cmidrule{2-9}
  & mean & 95\% CI & mean & 95\% CI & mean & 95\% CI & mean & 95\% CI \\ 
  \midrule
  NOIG & 0.50 & (0.49;0.50) & 0.58 & (0.56;0.60) & 3894 & (3685;4254) & 5575 & (5351;5742) \\[0.25em] 
  IG & 0.50 & (0.49;0.52) & 0.57 & (0.55;0.60) & 4223 & (3972;4591) & 5627 & (5288;5944) \\[0.25em] 
  BENCH & 0.50 & (0.48;0.52) & 0.57 & (0.55;0.60) & 4222 & (3973;4592) & 5627 & (5288;5944) \\[0.25em] 
  MNAR L & 0.49 & (0.47;0.52) & 0.56 & (0.53;0.58) & 4313 & (3850;4722) & 5689 & (5225;6135) \\[0.25em] 
  MNAR M & 0.49 & (0.47;0.52) & 0.56 & (0.53;0.59) & 4311 & (3863;4692) & 5675 & (5188;6143) \\[0.25em] 
  MNAR H & 0.49 & (0.47;0.52) & 0.56 & (0.53;0.59) & 4309 & (3875;4695) & 5672 & (5183;6156) \\ 
   \bottomrule
\end{tabular}
}
\end{table}
Across all scenarios, and within each intervention group, the QALYs mean estimates are similar both in terms of point estimates and HPD intervals, with slightly lower estimates under the three MNAR scenarios compared to the others. Total cost means show relatively larger differences with estimates under NOIG being considerably lower compared to those obtained under IG and BENCH, particularly in the control group. All three nonignorable departure scenarios (MNAR L, MNAR M, MNAR H) show total costs means that are higher compared to the estimates under ignorability or the benchmark nonignorable scenario (NOIG, IG, BENCH). In general, variations are larger in the control, which has higher proportions of noncompleters with respect to the intervention.

\section{Economic Evaluation}\label{evaluation}
We complete the analysis by assessing the cost-effectiveness of the new intervention with respect to the control, comparing the results under nonignorable MAR without any restrictions (NOIG), ignorable based on MAR restrictions (IG), benchmark nonignorable (BENCH) and the three alternative nonignorable departure scenarios (MNAR L, MNAR M, MNAR H). We specifically rely on the examination of the Cost-Effectiveness Plane~\citep[CEP;][]{Black} and the Cost-Effectiveness Acceptability Curve~\citep[CEAC;][]{VanHout} to summarise the economic~analysis. 

\begin{figure}[!h]
\centering
\subfloat[]{\includegraphics[scale=0.45]{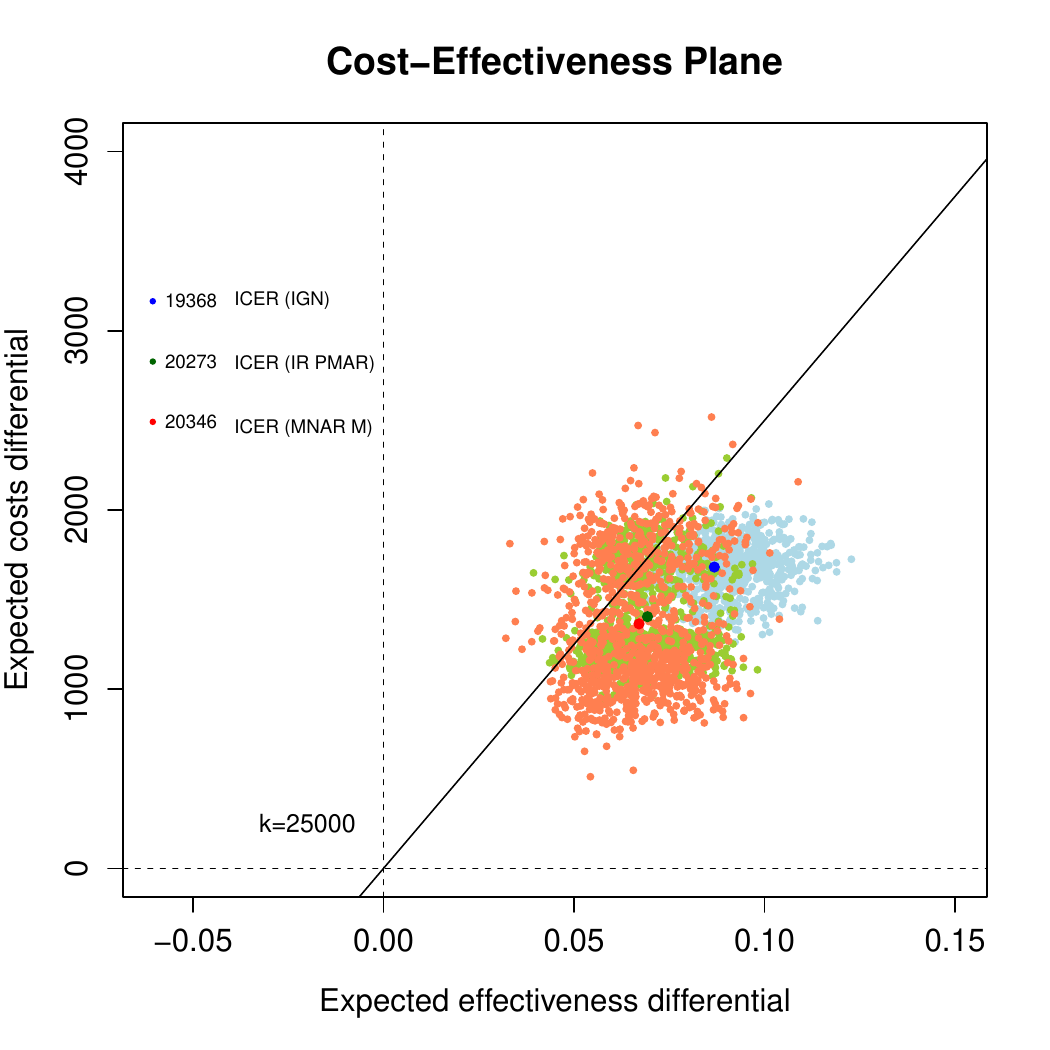}}
\subfloat[]{\includegraphics[scale=0.45]{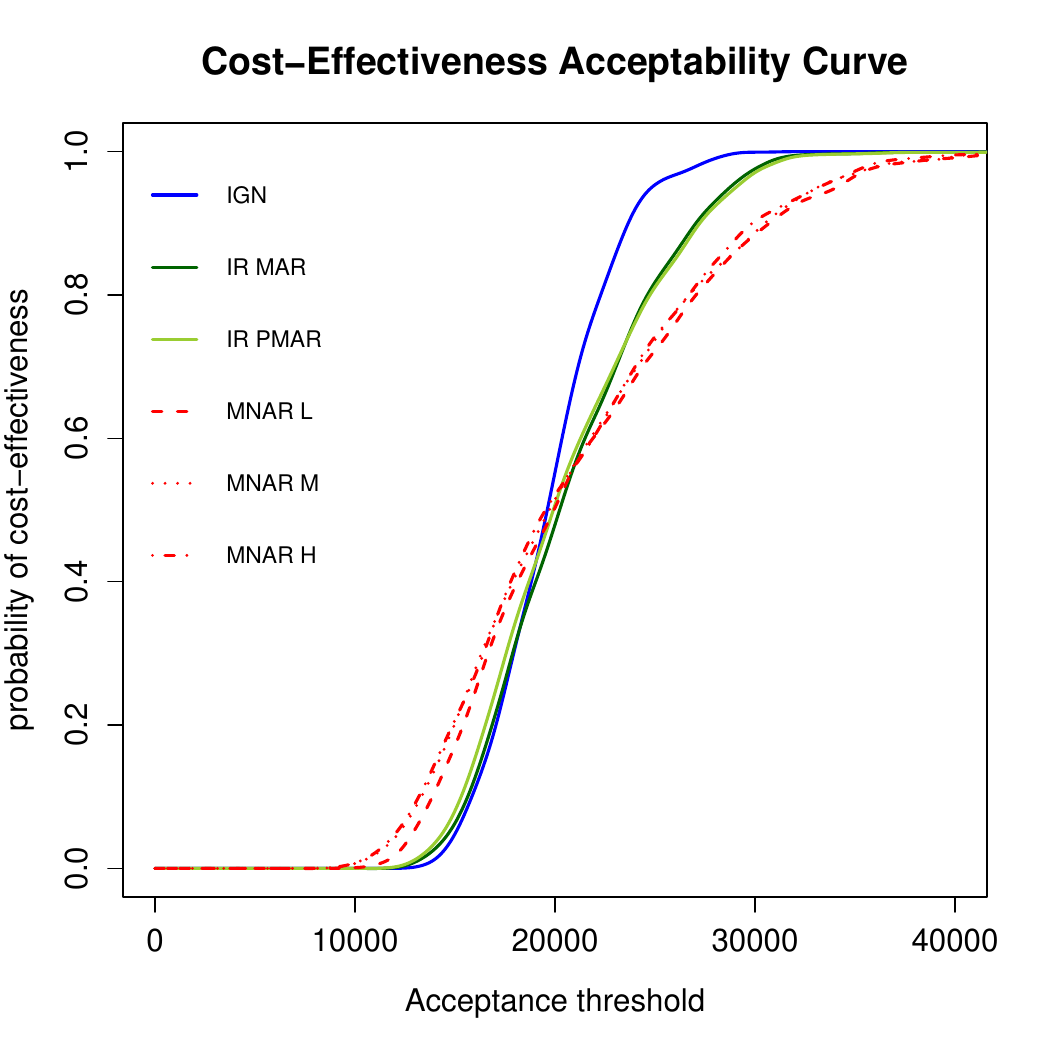}}
\caption{CEPs (panel a) and CEACs (panel b) associated with alternative missingness scenarios. In the CEPs, the ICERs based on the results from NOIG, BENCH and MNAR M are indicated with corresponding darker coloured dots, while the portion of the plane on the right-hand side of the straight line passing through the plot (evaluated at $k=\text{\pounds{}}25,000$) denotes the sustainability area. For the CEACs, in addition to the results under NOIG and IG and BENCH (solid lines), the probability values for the alternative MNAR scenarios are represented with different types of red dashed lines.}\label{CEAC2}
\end{figure}

The CEP (Figure~\ref{CEAC2}, panel a) is a graphical representation of the joint distribution for the population average effectiveness and costs increments between the two arms, indicated respectively as $\mu_{e2}-\mu_{e1}$ and $\mu_{c2}-\mu_{c1}$. We show the results only under three scenarios (light blue for NOIG, light green for BENCH and light red for MNAR M) for clarity and visualisation purposes. The slope of the straight line crossing the plane is the ``willingness to pay'' threshold (often indicated as $k$). This can be considered as the amount of budget the decision-maker is willing to spend to increase the health outcome of one unit and effectively is used to trade clinical benefits for money. Points lying below this straight line fall in the so-called sustainability area and suggest that the active intervention is more cost-effective than the control. In the graph, we also show the Incremental Cost-Effectiveness Ratio (ICER) computed under each scenario, as darker coloured dots. This is defined as 
$$\mbox{ICER}=\frac{\mbox{E}[\mu_{c2}-\mu_{c1}]}{\mbox{E}[\mu_{e2}-\mu_{e1}]},$$ 
and quantifies the cost per incremental unit of effectiveness. For all three scenarios almost all samples fall in the North-East quadrant and are associated with positive ICERs. This suggests that the intervention is likely to produce both QALY gains and cost savings. The ICER under MNAR M falls in the sustainability area and is the highest among the models compared, therefore indicating a slightly less positive cost-effective assessment for the new intervention with respect to NOIG and BENCH.

The CEAC (Figure~\ref{CEAC2}, panel b) is obtained by computing the proportion of points lying in the sustainability area upon varying the willingness to pay threshold $k$. Based on standard practice in routine analyses, we consider a range for $k$ up to \pounds{40,000} per QALY gained. The CEAC estimates the probability of cost-effectiveness, thus providing a simple summary of the uncertainty associated with the ``optimal'' decision-making suggested by the ICER. The results under NOIG, IG and BENCH are reported using blue, light green and dark green solid lines, respectively. In addition, the results derived under the three MNAR scenarios are reported using different types of red dashed lines. 
The CEACs under all scenarios show a similar trend up to $k=\pounds20000$ with a maximum probability of cost-effectiveness for the new intervention of about $0.5$. However, for $k$ between $\pounds20000$ and $\pounds30000$, results show notable differences between three sets of analyses: NOIG is associated with the steepest increment in the curve indicating the highest probability of cost-effectiveness; IG and BENCH show a moderate and almost identical rise of the curves; all three MNAR scenarios indicate the lowest increment in the cost-effectiveness probability, with relatively small differences between each other. Overall, the CEAC plot shows that results are sensitive to the assumptions about the missing values, which can lead to a considerable change in the output of the decision process and the cost-effectiveness~conclusions.

\section{Discussion}\label{discussion}
Missingness represents a threat to economic evaluations as, when dealing with partially-observed data, any analysis makes assumptions about the missing values that cannot be verified from the data at hand. Trial-based analyses are typically conducted on the observed utility and cost data from the completers in the study and may produce biased results, unless the completers are a random sample of all study participants which is very unlikely. A further concern is that routine analyses typically rely on standard models that ignore, or at best fail, to properly account for potentially important features in the data such as correlation, skewness, and the presence of structural~values.

In this paper, we have proposed an alternative approach for conducting nonparametric Bayesian inference under nonignorable missingness for a longitudinal bivariate outcome in health economic evaluations, which allows us to retain data from all partially-observed individuals in the analysis while also improving the fit to the observed utilities and costs compared to standard parametric models. The analysis of the PBS data shows the benefits of using our approach by showing a considerable impact of alternative MNAR assumptions on the final decision-making conclusions, suggesting a less cost-effective intervention compared with the results obtained under ignorability. 

We relied on the extrapolation factorisation, within a pattern mixture approach, and handled the sparsity of the data in most patterns using a working model to draw inferences about the distribution of the observed data while leaving the extrapolation distribution unidentified. Within our benchmark scenario we then identified the extrapolation distribution via identifying restrictions conditional on the dropout indicators $d^{min}$ and $d^{max}$ for the bivariate health economic outcome. This allows us to fully exploit the information from the observed utility and cost patterns to identify the conditional distributions for each type of outcome based on the corresponding dropout time. Next, we used sensitivity parameters to characterise the uncertainty about the missing data within each dropout pattern. Within a copula framework, alternative prior choices, calibrated using the observed data and different correlation values, were used to specify the sensitivity parameters and assess the robustness of the results across different scenarios.

A possible extension area for future work is to increase the flexibility of our approach by embedding more flexible parametric specifications within the Dirichlet process mixture for modelling the observed data distribution, which have only been explored within a fully parametric modelling framework~\citep{gabrio2020bayesian}. This, in turn, would allow a weakening of the model assumptions and likely further improve the fit of the model to the observed data, although it will likely increase the computational cost for implementing the model.

In conclusion, in this work we have presented a flexible Bayesian analytic framework that can: \textit{a)} jointly model costs and utilities using a flexible and longitudinal nonparametric approach; \textit{b)} identify the model under a benchmark nonignorable scenario based on outcome-specific dropout indicators from which the robustness of the results to a set of departures can be assessed via informative priors and sensitivity parameters. In order to increase the accessibility of the proposed methods, we are working on the development of an \texttt{R} package which allows to implement the proposed model in a relatively user-friendly way while also allowing a flexible specification for the missing data assumptions. 


\ack{Dr Michael J. Daniels was partially supported by the US NIH grant CA-183854.\\

\clearpage

\bibliographystyle{apa}
\bibliography{report}

\end{document}